\documentclass{article}

\usepackage{amssymb,amsfonts,amsmath}
\usepackage{cite,enumerate,float}
\usepackage{color}

\usepackage{physics}

\usepackage{mathrsfs}

\usepackage{tikz}
\usetikzlibrary{arrows}
\usetikzlibrary{arrows.meta}
\usetikzlibrary{positioning}
\usetikzlibrary{shapes,snakes}
\usetikzlibrary{fit}
\usetikzlibrary{decorations.pathmorphing,decorations.pathreplacing,decorations.markings}
\usetikzlibrary{calc}

\usepackage{ytableau}

\usepackage{xcolor}
\definecolor{burgundy}{rgb}{0.5, 0.0, 0.13}
\definecolor{olive}{rgb}{0.50, 0.50, 0.0}
\usepackage[linktocpage=true,colorlinks=true,linkcolor=burgundy,citecolor=black!20!blue,urlcolor=black!40!violet]{hyperref}

\usepackage[most]{tcolorbox}

\usepackage{pdflscape}
\usepackage{array, longtable}
\newcolumntype{C}{>{$}c<{$}}

\def\be{\begin{eqnarray}}
\def\ee{\end{eqnarray}}
\def\nn{\nonumber}

\def\p{\partial}

\definecolor{red}{rgb}{1,0,0}
\definecolor{orange}{rgb}{1,0.5,0}
\definecolor{violet}{rgb}{0.7,0,1}

\def\CH {{\cal H}}

\def\CN {{\cal N}}

\def\CV {{\cal V}}

\def\CH {{\cal H}}

\def\IC{\mathbb{C}}

\def\IN{\mathbb{N}}

\def\IR{{\mathbb{R}}}

\def\IZ{{\mathbb{Z}}}

\def\fg{\mathfrak{g}}

\def\fl{\mathfrak{l}}

\def\lm{\limits}
\def\nn{\nonumber}

\def\Yscale{0.1}

\DeclareSymbolFont{bbsymbol}{U}{bbold}{m}{n}
\DeclareMathSymbol{\bbzero}{\mathbin}{bbsymbol}{"30}
\DeclareMathSymbol{\bbone}{\mathbin}{bbsymbol}{"31}
\DeclareMathSymbol{\bbtwo}{\mathbin}{bbsymbol}{"32}
\DeclareMathSymbol{\bbthree}{\mathbin}{bbsymbol}{"33}
\DeclareMathSymbol{\bbfour}{\mathbin}{bbsymbol}{"34}
\DeclareMathSymbol{\bbfive}{\mathbin}{bbsymbol}{"35}
\DeclareMathSymbol{\bbsix}{\mathbin}{bbsymbol}{"36}
\DeclareMathSymbol{\bbseven}{\mathbin}{bbsymbol}{"37}
\DeclareMathSymbol{\bbeight}{\mathbin}{bbsymbol}{"38}
\DeclareMathSymbol{\bbnine}{\mathbin}{bbsymbol}{"39}

\newcommand\sqbox[1]{{
		\setbox0=\hbox{\mbox{$\Box$}}
		\setbox1=\hbox{\mbox{\raisebox{0.35ex}{\small #1}}}
		\mbox{\raisebox{-0.2ex}{\rlap{\hbox to \wd0{\hss{\box1}\hss}}\box0}}
}}
\newcommand\ssqbox[1]{{
		\setbox0=\hbox{\mbox{$\scriptstyle\Box$}}
		\setbox1=\hbox{\mbox{\raisebox{0.35ex}{\tiny #1}}}
		\mbox{\raisebox{-0.2ex}{\rlap{\hbox to \wd0{\hss{\box1}\hss}}\box0}}
}}

\tcbset{boxrule=0.6pt,colback=white!97!blue}
\hoffset= - 1.0in         % switch off for draft style
\begin{document}

\hfill MIPT/TH-03/24

\hfill ITEP/TH-04/24

\hfill IITP/TH-03/24

\vskip 1.5in
%\vskip 1cm
\begin{center}
	
{\bf\Large Simple Representations of BPS Algebras: the case of $Y(\widehat{\mathfrak{gl}}_2)$} \\
%	{\bf\Large{???Uglov-related/inspired??? representation of $Y(\widehat{\fg\fl}_2)$\\
%	???Smooth slices of toric Calabi-Yau 3-folds: BPS algebras, hooks and orthogonal polynomials\\
%	???Simple and Explicit Representations of BPS Algebras. \\
%	Smooth Quiver Varieties, Hooks and Orthogonal Polynomials}}
%	%\vskip 1cm
	\vskip 0.2in
	\renewcommand{\thefootnote}{\fnsymbol{footnote}}
	{Dmitry Galakhov$^{2,3,4,}$\footnote[2]{e-mail: galakhov@itep.ru},  Alexei Morozov$^{1,2,3,4,}$\footnote[3]{e-mail: morozov@itep.ru} and Nikita Tselousov$^{1,2,4,}$\footnote[4]{e-mail: tselousov.ns@phystech.edu}}
	\vskip 0.2in
	\renewcommand{\thefootnote}{\roman{footnote}}
	{\small{
			\textit{$^1$MIPT, 141701, Dolgoprudny, Russia}
			\vskip 0 cm
			\textit{$^2$NRC “Kurchatov Institute”, 123182, Moscow, Russia}
			\vskip 0 cm
			\textit{$^3$IITP RAS, 127051, Moscow, Russia}
			\vskip 0 cm
			\textit{$^4$ITEP, Moscow, Russia}
	}}
\end{center}

\vskip 0.2in
\baselineskip 16pt

\centerline{ABSTRACT}

\bigskip

{\footnotesize
	
    BPS algebras are the symmetries of a wide class of brane-inspired models.
    They are closely related to Yangians -- the peculiar and somewhat sophisticated limit of DIM algebras.
    Still they possess some simple and explicit representations.
    We explain here that for $Y(\widehat{\mathfrak{gl}}_r)$ these representations are related to Uglov polynomials, whose families are also labeled by natural $r$.
	They arise in the limit $\hbar\longrightarrow 0$ from Macdonald polynomials,
	and generalize the well-known Jack polynomials ($\beta$-deformation of Schur functions), associated with $r=1$.
	For $r=2$ they approximate Macdonald polynomials with the accuracy $O(\hbar^2)$,
	so that they are eigenfunctions of {\it two} immediately available commuting operators,
	arising from the $\hbar$-expansion of the first Macdonald Hamiltonian.
	These operators have a clear structure, which is easily generalizable, --
    what provides a technically simple way to build an explicit representation
	of Yangian $Y(\widehat{\mathfrak{gl}}_2)$, where $U^{(2)}$ are associated with the
	states $|\lambda\rangle$, parametrized by chess-colored Young diagrams.
	An interesting feature of this representation is that the odd time-variables
	$p_{2n+1}$ can be expressed through mutually commuting operators from Yangian,
	 however even time-variables $p_{2n}$ are inexpressible.
    Implications to higher $r$ become now straightforward, yet we describe them only in a sketchy way.

}

\bigskip

\bigskip

\tableofcontents

\ytableausetup{boxsize = 0.3em}
\section{Introduction}

%We should stress that this note follows up sequentially the program proposed in
This paper is a part of the program, proposed in
\cite{Galakhov:2023mak, Galakhov:2023gjs} for the study of Yangian algebras in terms of their
physically relevant representations.
Algebra $Y(\widehat{\fg\fl}_r)$ is thought \cite{Li:2020rij, Rapcak:2020ueh} to be a BPS algebra \cite{Harvey:1995fq} of D-brane states in type IIA string theory wrapping a toric Calabi-Yau resolution of singularity $(\IC^2/\IZ_r)\times\IC$ \cite{Douglas:1996sw}.
In particular, we will consider representations associated to Fock modules.
Those representations are expected to arise when branes are confined to a smooth 4-cycle.
The effective field theory is a quiver field theory, whose quiver is of Nakajima type \cite{2009arXiv0905.0686G} with enhanced supersymmetry.
This quiver resembles the Dynkin diagram of $\widehat{\fg\fl}_r$ (see Fig.~\ref{fig:QDglR}).
As well this quiver theory emerges as an effective description of non-commutative $U(1)$ instantons (Hilbert schemes) on $\IR^4/\IZ_r$\cite{ALE} via the McKay correspondence (see e.g. \cite{Cirafici:2012qc}).
In general, according to \cite{Galakhov:2023mak} for such types of smooth slices we expect the following \emph{phenomena}:
\begin{enumerate}
	\item The representation vectors are labeled by (some modification of) Young diagrams.
	And the Euler class of the tangent bundles to the the fixed points on the smooth quiver variety is given by a product of ``hooks'' in the corresponding diagram.
        \item There is an infinite sequence of Hamiltonian operators $p_k$ (usually referred as ``time''-variables) in this representation allowing one to rewrite the wave functions spanning the BPS Hilbert space as a system of orthogonal (Schur, super-Schur, Jack, Uglov, Macdonald) polynomials in times $p_k$.
	\item The algebra generator matrix coefficients (the Littlewood-Richardson coefficients for our polynomials) are rational functions of the equivariant weights unlike meromorphic square roots for more generic Macmahon representations (see e.g. \cite[eq. (3.18)]{Li:2020rij}).
	
\end{enumerate}

\definecolor{palette1}{rgb}{0.603922, 0.466667, 0.811765}
\definecolor{palette2}{rgb}{0.329412, 0.219608, 0.517647}
\definecolor{palette3}{rgb}{0.0156863, 0.282353, 0.333333}
\definecolor{palette4}{rgb}{0.631373, 0.211765, 0.439216}
\definecolor{palette5}{rgb}{0.92549, 0.254902, 0.462745}
\definecolor{palette6}{rgb}{1., 0.643137, 0.368627}
\definecolor{palette7}{rgb}{0.313725, 0.45098, 0.85098}
\begin{figure}[ht!]
	\begin{center}
		\begin{tikzpicture}
			\foreach \i in {0,1,...,5}
			{
				\begin{scope}[rotate = 51.4286*\i]
					\draw[postaction={decorate},decoration={markings, mark= at position 0.6 with {\arrow{stealth}}}] ([shift=(0:1.55)]0,0) arc (0:51.4286:1.55);
					\draw[postaction={decorate},decoration={markings, mark= at position 0.4 with {\arrowreversed{stealth}}}] ([shift=(0:1.45)]0,0) arc (0:51.4286:1.45);
				\end{scope}
			}
			\foreach \i in {6}
			{
				\begin{scope}[rotate = 51.4286*\i]
					\draw[dashed] ([shift=(0:1.55)]0,0) arc (0:51.4286:1.55);
					\draw[dashed] ([shift=(0:1.45)]0,0) arc (0:51.4286:1.45);
				\end{scope}
			}
			\foreach \i/\c/\x in {0/palette1/$\scriptstyle r$, 1/palette2/$\scriptstyle 1$, 2/palette3/$\scriptstyle 2$, 3/palette4/$\scriptstyle 3$, 4/palette6/$\scriptstyle 4$, 5/palette7/$\scriptstyle 5$, 6/palette5/$\scriptstyle 6$}
			{
				\begin{scope}[rotate=51.4286*\i]
					\draw[postaction={decorate},decoration={markings, mark= at position 0.7 with {\arrow{stealth}}}] (1.5,0.05) -- (2,0.05);
					\draw[postaction={decorate},decoration={markings, mark= at position 0.5 with {\arrow{stealth}}}] (2,-0.05) -- (1.5,-0.05);
					\begin{scope}[shift={(2,0)}]
						\draw[fill=white] (-0.08,-0.08) -- (-0.08,0.08) -- (0.08,0.08) -- (0.08,-0.08) -- cycle;
					\end{scope}
					\draw[postaction={decorate},decoration={markings, mark= at position 0.8 with {\arrow{stealth}}}] (1.5,0) to[out=150,in=0] (1.1,0.2) to[out=180,in=90] (0.9,0) to[out=270,in=180] (1.1,-0.2) to[out=0,in=210] (1.5,0);
					\draw[fill=\c] (1.5,0) circle (0.15);
					\node[white] at (1.5,0) {\x};
				\end{scope}
			}
			%%%%%%%%%%%%%%%%%%%
			%%%%%%%%%%%%%%%%%%%
			\begin{scope}[shift={(4.5,0)}]
				\foreach \i in {0,1,...,5}
				{
					\begin{scope}[rotate = 51.4286*\i]
						\draw[thick] ([shift=(0:1.5)]0,0) arc (0:51.4286:1.5);
					\end{scope}
				}
				\foreach \i in {6}
				{
					\begin{scope}[rotate = 51.4286*\i]
						\draw[dashed] ([shift=(0:1.5)]0,0) arc (0:51.4286:1.5);
					\end{scope}
				}
				\foreach \i/\c/\x in {0/palette1/$\scriptstyle r$, 1/palette2/$\scriptstyle 1$, 2/palette3/$\scriptstyle 2$, 3/palette4/$\scriptstyle 3$, 4/palette6/$\scriptstyle 4$, 5/palette7/$\scriptstyle 5$, 6/palette5/$\scriptstyle 6$}
				{
					\begin{scope}[rotate=51.4286*\i]
						\draw[fill=\c] (1.5,0) circle (0.15);
						\node[white] at (1.5,0) {\x};
					\end{scope}
				}
			\end{scope}
			%%%%%%%%%%%%%%%%%%%%
			%%%%%%%%%%%%%%%%%%%%
			\node at (9,0) {$\begin{array}{c}
				\begin{tikzpicture}[scale=0.25]
					\draw[-stealth] (-1,1) -- (12,1);
					\draw[-stealth] (-1,1) -- (-1,-12);
					\node[right] at (12,1) {$\scriptstyle x$};
					\node[right] at (-1,-12) {$\scriptstyle y$};
					\foreach \x/\y/\t/\c in {0/0/1/palette2, 0/-1/5/palette7, 0/-2/4/palette6, 0/-3/3/palette4, 0/-4/2/palette3, 0/-5/1/palette2, 0/-6/5/palette7, 0/-7/4/palette6, 0/-8/3/palette4, 0/-9/2/palette3, 0/-10/1/palette2, 1/0/2/palette3, 1/-1/1/palette2, 1/-2/5/palette7, 1/-3/4/palette6, 1/-4/3/palette4, 1/-5/2/palette3, 1/-6/1/palette2, 1/-7/5/palette7, 1/-8/4/palette6, 1/-9/3/palette4, 1/-10/2/palette3, 2/0/3/palette4, 2/-1/2/palette3, 2/-2/1/palette2, 2/-3/5/palette7, 2/-4/4/palette6, 2/-5/3/palette4, 2/-6/2/palette3, 2/-7/1/palette2, 2/-8/5/palette7, 2/-9/4/palette6, 2/-10/3/palette4, 3/0/4/palette6, 3/-1/3/palette4, 3/-2/2/palette3, 3/-3/1/palette2, 3/-4/5/palette7, 3/-5/4/palette6, 3/-6/3/palette4, 3/-7/2/palette3, 3/-8/1/palette2, 3/-9/5/palette7, 3/-10/4/palette6, 4/0/5/palette7, 4/-1/4/palette6, 4/-2/3/palette4, 4/-3/2/palette3, 4/-4/1/palette2, 4/-5/5/palette7, 4/-6/4/palette6, 4/-7/3/palette4, 4/-8/2/palette3, 4/-9/1/palette2, 4/-10/5/palette7, 5/0/1/palette2, 5/-1/5/palette7, 5/-2/4/palette6, 5/-3/3/palette4, 5/-4/2/palette3, 5/-5/1/palette2, 5/-6/5/palette7, 5/-7/4/palette6, 5/-8/3/palette4, 5/-9/2/palette3, 5/-10/1/palette2, 6/0/2/palette3, 6/-1/1/palette2, 6/-2/5/palette7, 6/-3/4/palette6, 6/-4/3/palette4, 6/-5/2/palette3, 6/-6/1/palette2, 6/-7/5/palette7, 6/-8/4/palette6, 6/-9/3/palette4, 6/-10/2/palette3, 7/0/3/palette4, 7/-1/2/palette3, 7/-2/1/palette2, 7/-3/5/palette7, 7/-4/4/palette6, 7/-5/3/palette4, 7/-6/2/palette3, 7/-7/1/palette2, 7/-8/5/palette7, 7/-9/4/palette6, 7/-10/3/palette4, 8/0/4/palette6, 8/-1/3/palette4, 8/-2/2/palette3, 8/-3/1/palette2, 8/-4/5/palette7, 8/-5/4/palette6, 8/-6/3/palette4, 8/-7/2/palette3, 8/-8/1/palette2, 8/-9/5/palette7, 8/-10/4/palette6, 9/0/5/palette7, 9/-1/4/palette6, 9/-2/3/palette4, 9/-3/2/palette3, 9/-4/1/palette2, 9/-5/5/palette7, 9/-6/4/palette6, 9/-7/3/palette4, 9/-8/2/palette3, 9/-9/1/palette2, 9/-10/5/palette7, 10/0/1/palette2, 10/-1/5/palette7, 10/-2/4/palette6, 10/-3/3/palette4, 10/-4/2/palette3, 10/-5/1/palette2, 10/-6/5/palette7, 10/-7/4/palette6, 10/-8/3/palette4, 10/-9/2/palette3, 10/-10/1/palette2}
					{
						\draw[fill=\c] (\x,\y) -- (\x+1,\y) -- (\x+1,\y-1) -- (\x,\y - 1) -- cycle;
						\node[white] at (\x+0.5,\y-0.5) {\tiny \t};
					}
				\end{tikzpicture}
				\end{array}$};
			%%%%%%%%%%%%%%%%%%%%%%%%%%%%%%%%%%%%%%%%%
			%%%%%%%%%%%%%%%%%%%%%%%%%%%%%%%%%%%%%%%%%
			\node at (0,-2.5) {(a)};
			\node at (4.5,-2.5) {(b)};
			\node at (9,-2.5) {(c)};
		\end{tikzpicture}
	\end{center}
	\caption{a) quiver diagram for a 4-cycle in $(\IC^2/\IZ_r)\times\IC$;  b) Dynkin diagram for $\widehat{\fg\fl}_r$; c) a sample of canonical coloring of the cell field for the Fock module of $Y(\widehat{\fg\fl}_5)$.}\label{fig:QDglR}
\end{figure}

This paper is at the intersection of two hot topics:
the constructive theory of affine Yangians $Y(\widehat G)$\footnote{
	Due to an appearance of affine Yangians (and their generalizations) as BPS algebras of D-brane systems on Calabi-Yau 3-folds this subject acquired quite an extensive study over recent years.
	The reader might overview a subdirection palette in this area only from just a few samples of literature sources listed here \cite{Galakhov:2023gjs, Li:2023zub, Bao:2024noj, Butson:2023eid, Negut:2022pka, Prochazka:2023zdb,Bao:2023kkh}
	}
and the theory of Uglov polynomials $U_\lambda^{(r)}\{p_k\}$ \cite{Uglov:1997ia}.
The former is a far-going generalization of  the theory of commuting cut-and-join operators \cite{Mironov:2009cj}
(which form a Cartan-like subalgebra of the Yangian -- in fact one of many \cite{Mironov:2020pcd,Mironov:2023zwi,Mironov:2023wga,schiffmann2012cherednik, Smirnov:2021cyf}), while the latter attract increasing interest
in more conventional representation theory \cite{Litvinov:2020zeq, Chistyakova:2021yyd, Kolyaskin:2022tqi}
and matrix models \cite{MishMyak}.
We explain that the first nontrivial $U^{(2)}$ provides a peculiar representation
of affine Yangian
$Y(\widehat{\fg\fl}_2)$, that is a bit rarer character of the mathematical physics literature compared to canonical $Y(\widehat{\fg\fl}_1)$.
This relation generalizes the well known Jack representation of $Y(\widehat{\fg\fl}_1)$ \cite{nakajima1996jack}
and admits a straightforward extension to arbitrary $r>2$.
However, there is a technical simplification for $r=2$ which can be used as a clue
in the construction, thus in this paper we concentrate on this particular case.
Lifting to arbitrary $r$ will be described elsewhere.

We begin from reminding the definition of Uglov polynomials in sec.\ref{sec:Uglov}
and describe immediate implications from Macdonald theory for $r=2$ in sec.\ref{sec:Macdonald Hamiltonian}.
The main lesson is the shape of the relevant operators and their difference from
conventional cut-and-join operators \cite{Mironov:2009cj}, which were basic in the case of $r=1$.
The main new feature is non-polynomiality -- lifting of cut-and-join operators
to $r>2$ converts them into sums over Schur functions of arbitrary degree, as it was suggested in \cite{Mironov:2019uaa}.
Our study proves that reduction of the sums to polynomials of finite degree
was actually accidental -- as  anticipated in \cite{Mironov:2019uaa}.
Before the detailed exposition of these ideas in sec.\ref{sec:Fock reps} and \ref{sec: time var reps} we remind the basics
of Yangian algebra in sec.\ref{sec:Yangian}
We conclude in sec.\ref{sec:conc} with a short summary.
Appendices contain additional comments on the definition of Uglov polynomials (\ref{sec:Uglov_def}),
on LMNS interpretation of Pieri-like rules for gluing and cutting from to Young diagrams (\ref{sec:LMNS})
and on embedding of the Yangian representations into those of DIM (Ding-Iohara-Miki) algebra (\ref{sec:DIM}).

\section{Uglov polynomials}\label{sec:Uglov}

Uglov polynomials are defined as an $\hbar \longrightarrow 0$ limit of Macdonald polynomials
\be
U^{(r)}_\lambda\{p\} = \lim_{\hbar \longrightarrow 0} M^{q,t}_\lambda\{p\}
\ee
with $q = \exp\left( \hbar + \frac{2\pi i}{r}\right), t = \exp\left( \beta \hbar + \frac{2 \pi i}{r}\right)$, and $\lambda$ being a Young diagram\footnote{Young diagrams labeling the vectors of the $Y(\widehat{\fg\fl}_r)$ Fock modules appear naturally as slices of 3d crystals \cite{Bao:2024noj}. 
For details see \cite{Noshita:2021ldl, Noshita:2021dgj, Galakhov:2021xum, Galakhov:2023mak}.}.
For $r=1$ we have $t=q^\beta$ and the limit provides just the Jack polynomials
\begin{equation}
    U_\lambda^{(1)}\{p\} = J_\lambda \{p\}.
\end{equation}
For $\beta=1$ all Uglov polynomials turn into Schur functions
\begin{equation}
    U_\lambda^{(r)}\{p\} \ \stackrel{\beta=1}{=}\  S_\lambda\{p\}.
\end{equation}

To illustrate the $r$-dependence we list just a few polynomials of lower degrees:
\begingroup
\renewcommand*{\arraystretch}{1.8}
\begin{equation}
	\begin{array}{c|ccccccc}
		\lambda \ \diagdown \ r & 1 \ ({\rm Jack}) & 2 & 3 & 4 & 5 & \ldots \\
		\hline
		\phantom.\begin{array}{c}\begin{tikzpicture}[scale=0.2]
				\foreach \x/\y/\z/\w in {0/-1/1/-1, 0/0/0/-1, 0/0/1/0, 1/0/1/-1}
				{
					\draw (\x,\y) -- (\z,\w);
				}
		\end{tikzpicture}\end{array} & p_1 & p_1 & p_1 & p_1 & p_1 & \\
		\hline
		\phantom.\begin{array}{c}\begin{tikzpicture}[scale=0.2]
				\foreach \x/\y/\z/\w in {0/-1/1/-1, 0/0/0/-1, 0/0/1/0, 1/-1/2/-1, 1/0/1/-1, 1/0/2/0, 2/0/2/-1}
				{
					\draw (\x,\y) -- (\z,\w);
				}
		\end{tikzpicture}\end{array} &
		\frac{p_1^2-p_2}{2} & \frac{p_1^2-p_2}{2} &\frac{p_1^2-p_2}{2} &\frac{p_1^2-p_2}{2} &\frac{p_1^2-p_2}{2} &  \\
		\phantom.\begin{array}{c}\begin{tikzpicture}[scale=0.2]
				\foreach \x/\y/\z/\w in {0/-2/1/-2, 0/-1/0/-2, 0/-1/1/-1, 0/0/0/-1, 0/0/1/0, 1/-1/1/-2, 1/0/1/-1}
				{
					\draw (\x,\y) -- (\z,\w);
				}
		\end{tikzpicture}\end{array} &
		\frac{\beta p_1^2+p_2}{2} &\frac{p_1^2+\beta p_2}{2} &\frac{p_1^2+p_2}{2} &\frac{p_1^2+p_2}{2} &  \frac{p_1^2+p_2}{2} & \\
		\hline
		\phantom.\begin{array}{c}\begin{tikzpicture}[scale=0.2]
				\foreach \x/\y/\z/\w in {0/-1/1/-1, 0/0/0/-1, 0/0/1/0, 1/-1/2/-1, 1/0/1/-1, 1/0/2/0, 2/-1/3/-1, 2/0/2/-1, 2/0/3/0, 3/0/3/-1}
				{
					\draw (\x,\y) -- (\z,\w);
				}
		\end{tikzpicture}\end{array} & \frac{p_1^3-3p_2p_1+2p_3}{6}&   \frac{p_1^3-3p_2p_1+2p_3}{6}& \frac{p_1^3-3p_2p_1+2p_3}{6}&
		\frac{p_1^3-3p_2p_1+2p_3}{6}&\frac{p_1^3-3p_2p_1+2p_3}{6}&
		\\
		\phantom.\begin{array}{c}\begin{tikzpicture}[scale=0.2]
				\foreach \x/\y/\z/\w in {0/-2/1/-2, 0/-1/0/-2, 0/-1/1/-1, 0/0/0/-1, 0/0/1/0, 1/-1/1/-2, 1/-1/2/-1, 1/0/1/-1, 1/0/2/0, 2/0/2/-1}
				{
					\draw (\x,\y) -- (\z,\w);
				}
		\end{tikzpicture}\end{array} & \frac{\beta p_1^3-(\beta-1)p_2p_1 - p_3}{2\beta+1} & \frac{p_1^3-p_3}{3} &  \frac{p_1^3-p_3}{3} &
		\frac{p_1^3-p_3}{3} &\frac{p_1^3-p_3}{3} &
		\\
		\phantom.\begin{array}{c}\begin{tikzpicture}[scale=0.2]
				\foreach \x/\y/\z/\w in {0/-3/1/-3, 0/-2/0/-3, 0/-2/1/-2, 0/-1/0/-2, 0/-1/1/-1, 0/0/0/-1, 0/0/1/0, 1/-2/1/-3, 1/-1/1/-2, 1/0/1/-1}
				{
					\draw (\x,\y) -- (\z,\w);
				}
		\end{tikzpicture}\end{array} & \frac{\beta^2p_1^3+3\beta p_2p_1+2p_3}{(\beta+1)(\beta+2)}& \frac{p_1^3+3\beta p_2p_1+2p_3}{3(\beta+1)}&
		\frac{p_1^3+3p_2p_1+2\beta p_3}{2(\beta+2)}&\frac{p_1^3+3p_2p_1+2p_3}{6}&\frac{p_1^3+3p_2p_1+2p_3}{6}&
		\\
		\hline
		\ldots &
	\end{array}
\end{equation}
\endgroup
These polynomials can be also build by the standard orthogonalization method,
reviewed in \cite{Mironov:2019uaa, Belavin:2012eg} (see also \cite{nakajima1996jack} for canonical notations), with respect to the canonical norm
\be
\langle p_{\mu},p_{\lambda}\rangle=\delta_{\mu,\lambda}\,z_{\lambda}\,\beta^{\left|\left\{i|\lambda_i=0\;{\rm mod}\;r\right\}\right|}\,.
\ee
We sketch this definition in Appendix \ref{sec:Uglov_def}.
For $|\lambda|<r$ the polynomials do not depend on $\beta$ and become ordinary Schur polynomials.

Despite the difference of the Uglov polynomials from Jacks ones does not look drastic, in fact, it is.
The usual formalism of cut-and-join operators is not immediately applicable and the question arises
what is its correct substitute.
It turns out that the Yangian formulation is generalizable, only $Y(\widehat{\fg\fl}_1)$ for Jacks
\cite{Morozov:2022ndt,MT} is substituted by $Y(\widehat{\fg\fl}_r)$.
Further lift to $Y(\widehat{\fg\fl}_{m|n})$, generalizing \cite{Galakhov:2023mak}, is also straightforward.
However, important is the first step beyond $\widehat{\fg\fl}_1$.

As we will see it involves a number of peculiarities --
which are somewhat in parallel to lifting Kac-Moody bosonization from ``Abelian'' $\widehat{\fg\fl}_1$ to
generic non-Abelian $\widehat{G}$ \cite{GMMOS}.
The number of free fields is expected to increase with the rank -- from $r$ for the simplest Kac representation \cite{kac1990infinite}
with central charge $k=1$ to ${\rm dim}_G$ in general \cite{GMMOS}.
In the description below the $r$ Kac fields are somehow packed into a single field
(a single set of time-variables), but already for $r=2$
we will see that odd times begin to differ from the even ones.
For higher $r$ these kind of problems are supposed to become more pronounced.

The possibility to build the representation in explicit way is provided by a special feature
of Uglov polynomials at $r=2$: Macdonald polynomials approximate them with the enhanced accuracy $O(h^2)$.
This means that Uglov polynomials are eigenfunctions not only of the $\hbar\longrightarrow 0$ limit
${\cal H}_0^{(r=2)}$ of the
Macdonald Hamiltonian, but also of the first $\hbar$ correction ${\cal H}_1^{(r=2)}$ to it.
This helps to understand the structure of the Yangian generators in the representation on Uglov polynomials.
They are based on the magnificent fact that the lowest cut-and-join operators \cite{MMN}, like Hamiltonians, are
often expanded in bilinear combinations of a {\it single}-hook Schur functions \cite{Mironov:2019uaa},
and these expansions for the lowest Cartan $\psi$-generators of Yangian
imply those for the lowest??? raising and lowering generators $e$ and $f$.
Once this is realized, all the rest are just technicalities.

\section{Uglov limit of the Macdonald Hamiltonian at $r=2$}\label{sec:Macdonald Hamiltonian}

The first Macdonald Hamiltonian in terms of time-variables has the following form:
\begin{equation}
    \hat{\mathcal{H}} = \oint \frac{dz}{z} \exp \left( \sum^{\infty}_{k=0} \frac{(1-t^{-k})}{k} \, p_k \, z^k\right) \exp \left( \sum^{\infty}_{k=0} \frac{(q^{k}-1)}{z^k} \, \frac{\partial}{\partial p_k}\right)
\end{equation}
\begin{equation}
    q = \exp\left( \hbar + \frac{2\pi i}{r}\right), \hspace{15mm} t = \exp\left( \beta \hbar + \frac{2 \pi i}{r}\right)
\end{equation}
We are interested in the first two terms in the $\hbar$-expansion:
\begin{equation}
    \hat{\mathcal{H}} = \hat{\mathcal{H}}_0 + \hbar \, \hat{\mathcal{H}}_1 + \ldots
\end{equation}
and due to the special property $r=2$ 
\begin{equation}
    M^{q,t}_\lambda\{p\} \ \stackrel{r=2}{=}\ U^{(2)}_\lambda\{p\} + \hbar \cdot 0 + O(\hbar^2)
\end{equation}
Uglov polynomials are the eigenfunctions of both $\hat{\mathcal{H}}_0$ and $\hat{\mathcal{H}}_1$.
Another intriguing fact is that both these operators at $r=2$ 
involve just the single-hook Schur functions $S_{(i|n-i)}\{p\}$ and their duals,
which are just the same functions, expressed through the  Heisenberg-dual operators:
\begin{equation}
    \hat{S}_R := S_R\left\{ p_{k}^{*}\right\}, \hspace{15mm} p_{k}^{*} := k\frac{\p}{\p p_k}
\end{equation}
Single-hook Young diagram with $n$ boxes and the first row of length $(n-i+1)$ in our notation has the form $(i|n-i) = [n-i+1, 1^{i-1}]$, i.e. the number of rows of this Young diagram is $i$. For example: $(2|2) = \begin{tikzpicture}[scale=0.2]
        \foreach \x/\y/\z/\w in {  0/-2/1/-2, 0/-1/0/-2, 0/-1/1/-1, 0/0/0/-1, 0/0/1/0,  1/-1/1/-2, 1/-1/2/-1, 1/0/1/-1, 1/0/2/0, 2/-1/3/-1, 2/0/2/-1, 2/0/3/0, 3/0/3/-1}
        {
	       \draw (\x,\y) -- (\z,\w);
        }
\end{tikzpicture}$.
In this notation these operators take the following simple form:
\begin{align}
    \begin{aligned}
        \hat{\mathcal{H}}_0 &= \sum_n (-)^n \sum_{i,j=1}^n  S_{(i|n-i)}\hat{S}_{(j|n-j)} = \sum_{n=1}^{\infty} (-)^n \mathcal{P}_{0,n} \widehat{\mathcal{P}}_{0,n}  = \\
&= -p_1\frac{\p}{\p p_1} + p_1^2\frac{\p^2}{\p p_1^2}
- \frac{p_3+2p_1^3}{9}\left(3\frac{\p}{\p p_3} + 2\frac{\p^3}{\p p_1^3}\right)
+ \frac{2p_3p_1+p_1^4}{9}\left(6\frac{\p^2}{\p p_3\p p_1}+ \frac{\p^4}{\p p_1^4}\right) + \ldots
    \end{aligned}
\end{align}

The next expansion term for $r=2$ is
\begin{align}
    \begin{aligned}
        \hat{\mathcal{H}}_1 &= \sum_{n=1}^{\infty} (-)^n\Big[ \mathcal{P}_{0,n} \widehat{\mathcal{P}}_{1,n} + \beta \, \mathcal{P}_{1,n} \widehat{\mathcal{P}}_{0,n}  + (1-\beta) \, n \, \mathcal{P}_{0,n} \widehat{\mathcal{P}}_{0,n}   \Big] = \\
        &= (\beta -1) p_1 \frac{\partial}{\partial p_1} + \left(\beta  p_2-2 (\beta -1) p_1^2\right) \frac{\partial^2}{\partial p_1^2}+2 p_1^2 \frac{\partial}{\partial p_2} + \ldots
    \end{aligned}
\end{align}
with 
\begin{equation}
    \mathcal{P}_{0,n} = \sum_{i=1}^n S_{(i|n-i)} \hspace{15mm}   \mathcal{P}_{1,n} =  \sum_{i=1}^n (n+1-2i) \, S_{(i|n-i)}
\end{equation}
For $r=2$ Uglov polynomials are eigenfunctions of both these operators:
\begin{equation}
\label{Ham 0 and 1}
    \hat{\mathcal{H}}_0 \, U^{(2)}_{\lambda} = \mathcal{E}_{\lambda,0} \, U^{(2)}_{\lambda}, \hspace{15mm} \hat{\mathcal{H}}_1 \, U^{(2)}_{\lambda} = \mathcal{E}_{\lambda,1} \, U^{(2)}_{\lambda}
\end{equation}
\begin{equation}
    \mathcal{E}_{\lambda,0} = n^{(2)}_{\lambda} - n^{(1)}_{\lambda}, \hspace{15mm} \mathcal{E}_{\lambda,1} = 2  \left( \kappa_{\lambda}^{(2)} - \kappa_{\lambda}^{(1)} \right) +(1-\beta)  \left(n^{(2)}_{\lambda} - n^{(1)}_{\lambda} \right) 
\end{equation}
where 
\begin{equation}
    n_{\lambda}^{(a)} = \sum_{\ssqbox{$a$} \in \lambda} 1, \hspace{15mm}\kappa_{\lambda}^{(a)} = \sum_{\ssqbox{$a$} \in \lambda} x_{\ssqbox{$a$}} - \beta y_{\ssqbox{$a$}}
\end{equation}
Here letter $(a)$ plays a role of the color of the box as depicted in Fig.\ref{fig:QDglR}. The color of a box that completely fixed by its own coordinates $(i,j)$:
\begin{equation}
    a = 1+ \Big[ (i+j)\mod 2 \Big]
\end{equation}
The above operators \eqref{Ham 0 and 1} will play the role of building blocks for the lowest Cartan operators $\hat{\psi}^{(a)}_0$ and $\hat{\psi}^{(a)}_1$ of the Yangian.
Given these spectacular expressions,
it is easy to guess that the first raising and lowering operators will differ
by just a shift of the diagram size:
\be
\hat e_0 \approx  \sum_n (-)^n \sum_{i,j=1}^n  S_{(i|n+1-i)}\hat{S}_{(j|n-j)} \nn \\
\hat f_0 \approx  \sum_n (-)^n \sum_{i,j=1}^n  S_{(i|n-i)}\hat{S}_{(j|n+1-j)}
\ee
%\be
%???
%\hat e_1^{(1)} + \hat e_1^{(2)} =\sum_a ap_{a+1}\frac{\p}{\p p_a} +
%(\beta-1)\left( p_1^2\frac{\p}{\p p_1} + \frac{2}{3} (p_3-p_1^3)\frac{\p^2}{\p p_1^2}
%+ \frac{1}{9} p_1(p_3-p_1^3)\Big(\frac{\p}{\p p_3}-4\frac{\p}{\p p_1^3}\Big)
%+ \right. \nn \\ \left.
%+ \frac{4}{5}(p_5-p_1^5)\frac{\p^2}{\p p_3p_1}
%- \frac{2}{9}p_1^2(p_3-p_1^3) \Big(4\frac{\p^2}{\p p_3\p p_1}-\frac{\p^4}{\p p_1^4}\Big)
%+ \ldots \right)
%\ee
and so on.

For $r>2$ one can expect $r$ different combinations like
$\sum_i   \omega_r^{ki} S_{(i|n-i)}$
with various $k$ for the $2r$-th root of unity $\omega_r=e^{\pi i/r}$.
Also, as explained in Appendix \ref{sec:Uglov_def},  more adequate for $r>1$ can be a version of  conjugation with additional factors of $\beta$.
We do not go into these considerations in the main body of the text.

For arbitrary $r$ Uglov polynomials are eigenfunctions of the $\beta$-independent operator

\begin{equation}
\begin{aligned}
\hat{\mathcal{H}}_0^{(r)} =
%???\left(\sin\frac{\pi}{r}\right)^2 \left\{
p_1\hat p_1
- \left(\sin\frac{\pi}{r}\cdot p_1^2-i \cos\frac{\pi}{r}\cdot  p_2\right)
\left(\sin\frac{\pi}{r}\cdot \hat p_1^2-i \cos\frac{\pi}{r}\cdot \hat p_2\right)
%+ \left(\sin^2\frac{\pi}{r}\cdot \hat p_1^3-3i \sin\frac{\pi}{r} \cos\frac{\pi}{r}\cdot \hat p_2\hat p_1\right)
+  \ldots
%\right\}
= \\
=  S_{\begin{tikzpicture}[scale=\Yscale]
		\foreach \x/\y/\z/\w in {0/-1/1/-1, 0/0/0/-1, 0/0/1/0, 1/0/1/-1}
		{
			\draw (\x,\y) -- (\z,\w);
		}
\end{tikzpicture}}\hat{S}_{\begin{tikzpicture}[scale=\Yscale]
\foreach \x/\y/\z/\w in {0/-1/1/-1, 0/0/0/-1, 0/0/1/0, 1/0/1/-1}
{
	\draw (\x,\y) -- (\z,\w);
}
\end{tikzpicture}}
+ \left(e^{\frac{\pi i}{r}}S_{\begin{tikzpicture}[scale=\Yscale]
		\foreach \x/\y/\z/\w in {0/-2/1/-2, 0/-1/0/-2, 0/-1/1/-1, 0/0/0/-1, 0/0/1/0, 1/-1/1/-2, 1/0/1/-1}
		{
			\draw (\x,\y) -- (\z,\w);
		}
\end{tikzpicture}} -e^{-\frac{\pi i}{r}}S_{\begin{tikzpicture}[scale=\Yscale]
\foreach \x/\y/\z/\w in {0/-1/1/-1, 0/0/0/-1, 0/0/1/0, 1/-1/2/-1, 1/0/1/-1, 1/0/2/0, 2/0/2/-1}
{
	\draw (\x,\y) -- (\z,\w);
}
\end{tikzpicture}}\right)
\left(e^{\frac{\pi i}{r}}\hat{S}_{\begin{tikzpicture}[scale=\Yscale]
		\foreach \x/\y/\z/\w in {0/-2/1/-2, 0/-1/0/-2, 0/-1/1/-1, 0/0/0/-1, 0/0/1/0, 1/-1/1/-2, 1/0/1/-1}
		{
			\draw (\x,\y) -- (\z,\w);
		}
\end{tikzpicture}}-e^{-\frac{\pi i}{r}}\hat{S}_{\begin{tikzpicture}[scale=\Yscale]
\foreach \x/\y/\z/\w in {0/-1/1/-1, 0/0/0/-1, 0/0/1/0, 1/-1/2/-1, 1/0/1/-1, 1/0/2/0, 2/0/2/-1}
{
	\draw (\x,\y) -- (\z,\w);
}
\end{tikzpicture}}\right) + \\
+ \left(e^{\frac{2\pi i}{r}}S_{\begin{tikzpicture}[scale=\Yscale]
		\foreach \x/\y/\z/\w in {0/-3/1/-3, 0/-2/0/-3, 0/-2/1/-2, 0/-1/0/-2, 0/-1/1/-1, 0/0/0/-1, 0/0/1/0, 1/-2/1/-3, 1/-1/1/-2, 1/0/1/-1}
		{
			\draw (\x,\y) -- (\z,\w);
		}
\end{tikzpicture}} -S_{\begin{tikzpicture}[scale=\Yscale]
\foreach \x/\y/\z/\w in {0/-2/1/-2, 0/-1/0/-2, 0/-1/1/-1, 0/0/0/-1, 0/0/1/0, 1/-1/1/-2, 1/-1/2/-1, 1/0/1/-1, 1/0/2/0, 2/0/2/-1}
{
	\draw (\x,\y) -- (\z,\w);
}
\end{tikzpicture}}+e^{-\frac{2\pi i}{r}}S_{\begin{tikzpicture}[scale=\Yscale]
\foreach \x/\y/\z/\w in {0/-1/1/-1, 0/0/0/-1, 0/0/1/0, 1/-1/2/-1, 1/0/1/-1, 1/0/2/0, 2/-1/3/-1, 2/0/2/-1, 2/0/3/0, 3/0/3/-1}
{
	\draw (\x,\y) -- (\z,\w);
}
\end{tikzpicture}}\right)
\left(e^{\frac{2\pi i}{r}}\hat{S}_{\begin{tikzpicture}[scale=\Yscale]
		\foreach \x/\y/\z/\w in {0/-3/1/-3, 0/-2/0/-3, 0/-2/1/-2, 0/-1/0/-2, 0/-1/1/-1, 0/0/0/-1, 0/0/1/0, 1/-2/1/-3, 1/-1/1/-2, 1/0/1/-1}
		{
			\draw (\x,\y) -- (\z,\w);
		}
\end{tikzpicture}} -\hat{S}_{\begin{tikzpicture}[scale=\Yscale]
\foreach \x/\y/\z/\w in {0/-2/1/-2, 0/-1/0/-2, 0/-1/1/-1, 0/0/0/-1, 0/0/1/0, 1/-1/1/-2, 1/-1/2/-1, 1/0/1/-1, 1/0/2/0, 2/0/2/-1}
{
	\draw (\x,\y) -- (\z,\w);
}
\end{tikzpicture}}+e^{-\frac{2\pi i}{r}}\hat{S}_{\begin{tikzpicture}[scale=\Yscale]
\foreach \x/\y/\z/\w in {0/-1/1/-1, 0/0/0/-1, 0/0/1/0, 1/-1/2/-1, 1/0/1/-1, 1/0/2/0, 2/-1/3/-1, 2/0/2/-1, 2/0/3/0, 3/0/3/-1}
{
	\draw (\x,\y) -- (\z,\w);
}
\end{tikzpicture}}\right)
+ \\
 + \left(e^{\frac{3\pi i}{r}}S_{\begin{tikzpicture}[scale=\Yscale]
 		\foreach \x/\y/\z/\w in {0/-4/1/-4, 0/-3/0/-4, 0/-3/1/-3, 0/-2/0/-3, 0/-2/1/-2, 0/-1/0/-2, 0/-1/1/-1, 0/0/0/-1, 0/0/1/0, 1/-3/1/-4, 1/-2/1/-3, 1/-1/1/-2, 1/0/1/-1}
 		{
 			\draw (\x,\y) -- (\z,\w);
 		}
 \end{tikzpicture}} -e^{\frac{\pi i}{r}}S_{\begin{tikzpicture}[scale=\Yscale]
 \foreach \x/\y/\z/\w in {0/-3/1/-3, 0/-2/0/-3, 0/-2/1/-2, 0/-1/0/-2, 0/-1/1/-1, 0/0/0/-1, 0/0/1/0, 1/-2/1/-3, 1/-1/1/-2, 1/-1/2/-1, 1/0/1/-1, 1/0/2/0, 2/0/2/-1}
 {
 	\draw (\x,\y) -- (\z,\w);
 }
\end{tikzpicture}} + e^{-\frac{\pi i}{r}}S_{\begin{tikzpicture}[scale=\Yscale]
\foreach \x/\y/\z/\w in {0/-2/1/-2, 0/-1/0/-2, 0/-1/1/-1, 0/0/0/-1, 0/0/1/0, 1/-1/1/-2, 1/-1/2/-1, 1/0/1/-1, 1/0/2/0, 2/-1/3/-1, 2/0/2/-1, 2/0/3/0, 3/0/3/-1}
{
	\draw (\x,\y) -- (\z,\w);
}
\end{tikzpicture}}
-  e^{-\frac{3\pi i}{r}}S_{\begin{tikzpicture}[scale=\Yscale]
		\foreach \x/\y/\z/\w in {0/-1/1/-1, 0/0/0/-1, 0/0/1/0, 1/-1/2/-1, 1/0/1/-1, 1/0/2/0, 2/-1/3/-1, 2/0/2/-1, 2/0/3/0, 3/-1/4/-1, 3/0/3/-1, 3/0/4/0, 4/0/4/-1}
		{
			\draw (\x,\y) -- (\z,\w);
		}
\end{tikzpicture}}\right)
\left(e^{\frac{3\pi i}{r}}\hat{S}_{\begin{tikzpicture}[scale=\Yscale]
		\foreach \x/\y/\z/\w in {0/-4/1/-4, 0/-3/0/-4, 0/-3/1/-3, 0/-2/0/-3, 0/-2/1/-2, 0/-1/0/-2, 0/-1/1/-1, 0/0/0/-1, 0/0/1/0, 1/-3/1/-4, 1/-2/1/-3, 1/-1/1/-2, 1/0/1/-1}
		{
			\draw (\x,\y) -- (\z,\w);
		}
\end{tikzpicture}} -e^{\frac{\pi i}{r}}\hat{S}_{\begin{tikzpicture}[scale=\Yscale]
\foreach \x/\y/\z/\w in {0/-3/1/-3, 0/-2/0/-3, 0/-2/1/-2, 0/-1/0/-2, 0/-1/1/-1, 0/0/0/-1, 0/0/1/0, 1/-2/1/-3, 1/-1/1/-2, 1/-1/2/-1, 1/0/1/-1, 1/0/2/0, 2/0/2/-1}
{
	\draw (\x,\y) -- (\z,\w);
}
\end{tikzpicture}} + e^{-\frac{\pi i}{r}}\hat{S}_{\begin{tikzpicture}[scale=\Yscale]
\foreach \x/\y/\z/\w in {0/-2/1/-2, 0/-1/0/-2, 0/-1/1/-1, 0/0/0/-1, 0/0/1/0, 1/-1/1/-2, 1/-1/2/-1, 1/0/1/-1, 1/0/2/0, 2/-1/3/-1, 2/0/2/-1, 2/0/3/0, 3/0/3/-1}
{
	\draw (\x,\y) -- (\z,\w);
}
\end{tikzpicture}}
-  e^{-\frac{3\pi i}{r}}\hat{S}_{\begin{tikzpicture}[scale=\Yscale]
		\foreach \x/\y/\z/\w in {0/-1/1/-1, 0/0/0/-1, 0/0/1/0, 1/-1/2/-1, 1/0/1/-1, 1/0/2/0, 2/-1/3/-1, 2/0/2/-1, 2/0/3/0, 3/-1/4/-1, 3/0/3/-1, 3/0/4/0, 4/0/4/-1}
		{
			\draw (\x,\y) -- (\z,\w);
		}
\end{tikzpicture}}\right) + \\
+ \left(e^{\frac{4\pi i}{r}}S_{\begin{tikzpicture}[scale=\Yscale]
		\foreach \x/\y/\z/\w in {0/-5/1/-5, 0/-4/0/-5, 0/-4/1/-4, 0/-3/0/-4, 0/-3/1/-3, 0/-2/0/-3, 0/-2/1/-2, 0/-1/0/-2, 0/-1/1/-1, 0/0/0/-1, 0/0/1/0, 1/-4/1/-5, 1/-3/1/-4, 1/-2/1/-3, 1/-1/1/-2, 1/0/1/-1}
		{
			\draw (\x,\y) -- (\z,\w);
		}
\end{tikzpicture}} -e^{\frac{2\pi i}{r}}S_{\begin{tikzpicture}[scale=\Yscale]
\foreach \x/\y/\z/\w in {0/-4/1/-4, 0/-3/0/-4, 0/-3/1/-3, 0/-2/0/-3, 0/-2/1/-2, 0/-1/0/-2, 0/-1/1/-1, 0/0/0/-1, 0/0/1/0, 1/-3/1/-4, 1/-2/1/-3, 1/-1/1/-2, 1/-1/2/-1, 1/0/1/-1, 1/0/2/0, 2/0/2/-1}
{
	\draw (\x,\y) -- (\z,\w);
}
\end{tikzpicture}} +S_{\begin{tikzpicture}[scale=\Yscale]
\foreach \x/\y/\z/\w in {0/-3/1/-3, 0/-2/0/-3, 0/-2/1/-2, 0/-1/0/-2, 0/-1/1/-1, 0/0/0/-1, 0/0/1/0, 1/-2/1/-3, 1/-1/1/-2, 1/-1/2/-1, 1/0/1/-1, 1/0/2/0, 2/-1/3/-1, 2/0/2/-1, 2/0/3/0, 3/0/3/-1}
{
	\draw (\x,\y) -- (\z,\w);
}
\end{tikzpicture}}- e^{-\frac{2\pi i}{r}}S_{\begin{tikzpicture}[scale=\Yscale]
\foreach \x/\y/\z/\w in {0/-2/1/-2, 0/-1/0/-2, 0/-1/1/-1, 0/0/0/-1, 0/0/1/0, 1/-1/1/-2, 1/-1/2/-1, 1/0/1/-1, 1/0/2/0, 2/-1/3/-1, 2/0/2/-1, 2/0/3/0, 3/-1/4/-1, 3/0/3/-1, 3/0/4/0, 4/0/4/-1}
{
	\draw (\x,\y) -- (\z,\w);
}
\end{tikzpicture}}
+  e^{-\frac{4\pi i}{r}}S_{\begin{tikzpicture}[scale=\Yscale]
		\foreach \x/\y/\z/\w in {0/-1/1/-1, 0/0/0/-1, 0/0/1/0, 1/-1/2/-1, 1/0/1/-1, 1/0/2/0, 2/-1/3/-1, 2/0/2/-1, 2/0/3/0, 3/-1/4/-1, 3/0/3/-1, 3/0/4/0, 4/-1/5/-1, 4/0/4/-1, 4/0/5/0, 5/0/5/-1}
		{
			\draw (\x,\y) -- (\z,\w);
		}
\end{tikzpicture}}\right)\times \\ \times
\left(e^{\frac{4\pi i}{r}}\hat{S}_{\begin{tikzpicture}[scale=\Yscale]
		\foreach \x/\y/\z/\w in {0/-5/1/-5, 0/-4/0/-5, 0/-4/1/-4, 0/-3/0/-4, 0/-3/1/-3, 0/-2/0/-3, 0/-2/1/-2, 0/-1/0/-2, 0/-1/1/-1, 0/0/0/-1, 0/0/1/0, 1/-4/1/-5, 1/-3/1/-4, 1/-2/1/-3, 1/-1/1/-2, 1/0/1/-1}
		{
			\draw (\x,\y) -- (\z,\w);
		}
\end{tikzpicture}} -e^{\frac{2\pi i}{r}}\hat{S}_{\begin{tikzpicture}[scale=\Yscale]
\foreach \x/\y/\z/\w in {0/-4/1/-4, 0/-3/0/-4, 0/-3/1/-3, 0/-2/0/-3, 0/-2/1/-2, 0/-1/0/-2, 0/-1/1/-1, 0/0/0/-1, 0/0/1/0, 1/-3/1/-4, 1/-2/1/-3, 1/-1/1/-2, 1/-1/2/-1, 1/0/1/-1, 1/0/2/0, 2/0/2/-1}
{
	\draw (\x,\y) -- (\z,\w);
}
\end{tikzpicture}} +\hat{S}_{\begin{tikzpicture}[scale=\Yscale]
\foreach \x/\y/\z/\w in {0/-3/1/-3, 0/-2/0/-3, 0/-2/1/-2, 0/-1/0/-2, 0/-1/1/-1, 0/0/0/-1, 0/0/1/0, 1/-2/1/-3, 1/-1/1/-2, 1/-1/2/-1, 1/0/1/-1, 1/0/2/0, 2/-1/3/-1, 2/0/2/-1, 2/0/3/0, 3/0/3/-1}
{
	\draw (\x,\y) -- (\z,\w);
}
\end{tikzpicture}}- e^{-\frac{2\pi i}{r}}\hat{S}_{\begin{tikzpicture}[scale=\Yscale]
\foreach \x/\y/\z/\w in {0/-2/1/-2, 0/-1/0/-2, 0/-1/1/-1, 0/0/0/-1, 0/0/1/0, 1/-1/1/-2, 1/-1/2/-1, 1/0/1/-1, 1/0/2/0, 2/-1/3/-1, 2/0/2/-1, 2/0/3/0, 3/-1/4/-1, 3/0/3/-1, 3/0/4/0, 4/0/4/-1}
{
	\draw (\x,\y) -- (\z,\w);
}
\end{tikzpicture}}
+  e^{-\frac{4\pi i}{r}}\hat{S}_{\begin{tikzpicture}[scale=\Yscale]
		\foreach \x/\y/\z/\w in {0/-1/1/-1, 0/0/0/-1, 0/0/1/0, 1/-1/2/-1, 1/0/1/-1, 1/0/2/0, 2/-1/3/-1, 2/0/2/-1, 2/0/3/0, 3/-1/4/-1, 3/0/3/-1, 3/0/4/0, 4/-1/5/-1, 4/0/4/-1, 4/0/5/0, 5/0/5/-1}
		{
			\draw (\x,\y) -- (\z,\w);
		}
\end{tikzpicture}}\right)
+ \ldots = \\
= \sum_{n=1}^\infty \sum_{i=1}^n  (-)^{i+j} e^{\frac{2\pi \sqrt{-1}}{r}(i+j-n-1)}S_{(i|n-i)} \hat{S}_{(j|n-j)}\,.
\end{aligned}
\end{equation}
It plays the role of the first Cartan operators $\hat{\psi}_0^{(a)}$. Actually, for different colors $a = 1,2$ these operators and their eigenvalues differ minimally.
At least for $r=2$ and $\beta=1$ the higher $\psi_k$ are also single-hook sums,
and multi-hook cut-and-join operators appear to be products of the generators $\hat{\psi}_k^{(a)}$, i.e. belong to the universal enveloping of the Yangian.

Uglov polynomials have a definite grading for all $r$--  but, as we will see,
the grading operator
\be
\hat W_{\begin{ytableau}  \  \end{ytableau}} = \sum_{a=1}^{\infty} a p_a \frac{\p}{\p p_a} = \sum_{n=1}^\infty \sum_{i,j=1}^n (-)^{i+j} S_{(i|n-i)}\hat{S}_{(j,n-j)}
\ee
is no longer an element of the Yangian algebra at $\beta\neq 1$, however for $\beta=1$ case it is.
Moreover, already the next standard cut-and-join operator
\be
\hat W_{\begin{ytableau}  \ & \  \end{ytableau}} = \sum_{a,b=1}^{\infty} \left(  ab p_{a+b}\frac{\p^2}{\p p_a\p p_b} + \beta (a+b) p_ap_b \frac{\p}{\p p_{a+b}}\right)
+ (1-\beta) \sum_{a=1}^{\infty} a(a-1) p_a \frac{\p}{\p p_a}
\ee
does not preserve Uglov polynomials for $r>1$.
Thus it requires some other substitute -- like $H_0^{(r)}$, which, is single-hook, but,
unlike $\hat W_{\begin{ytableau}  \ & \  \end{ytableau}}$ is not adjusted to imply numerous cancellations,
and is going to be a series in all powers of time-variables.

\section{Yangian $Y(\widehat{\mathfrak{gl}}_2)$: commutation relations} \label{sec:Yangian}
The algebra $Y(\widehat{\mathfrak{gl}}_2)$ has a Chevalley basis of generator families $\hat{e}^{(a)}_{n}, \hat{\psi}^{(a)}_{n}, \hat{f}^{(a)}_{n}$, $a=1,2, n = 0,1,2, \ldots$ that can be collected into generating series:
\begin{equation}
	\hat{e}^{(a)}(z) = \sum^{\infty}_{n = 0} \frac{\hat{e}^{(a)}_n}{z^{n+1}}, \hspace{10mm} \hat{\psi}^{(a)}(z) =-1 + \sum^{\infty}_{n = 0} \frac{\hat{\psi}^{(a)}_n}{z^{n+1}}, \hspace{10mm}
	\hat{f}^{(a)}(z) = \sum^{\infty}_{n = 0} \frac{\hat{f}^{(a)}_n}{z^{n+1}}
\end{equation}
Commutational relations have a simple form in terms of generating series:
\begin{align}
	\hat{\psi}^{(a)}(z) \, \hat{\psi}^{(b)}(w) &\sim  \hat{\psi}^{(b)}(w) \, \hat{\psi}^{(a)}(z) \\
	\Big[ \hat{e}^{(a)}(z), \hat{f}^{(b)}(w) \Big] &\sim \delta_{a,b} \frac{\hat{\psi}^{(a)}(z) - \hat{\psi}^{(a)}(w)}{z-w}
\end{align}
\begin{align}
	\hat{e}^{(a)}(z) \, \hat{e}^{(b)}(w) &\sim \varphi^{(a,b)}(z-w) \, \hat{e}^{(b)}(w) \, \hat{e}^{(a)}(z) \\
	\hat{f}^{(a)}(z) \, \hat{f}^{(b)}(w) &\sim \varphi^{(a,b)}(w-z) \, \hat{f}^{(b)}(w) \, \hat{f}^{(a)}(z)
\end{align}
\begin{align}
	\hat{\psi}^{(a)}(z) \, \hat{e}^{(b)}(w) &\sim \varphi^{(a,b)}(z-w) \, \hat{e}^{(b)}(w) \, \hat{\psi}^{(a)}(z) \\
	\hat{\psi}^{(a)}(z) \, \hat{f}^{(b)}(w) &\sim \varphi^{(a,b)}(w-z) \, \hat{f}^{(b)}(w) \, \hat{\psi}^{(a)}(z)
\end{align}

Where the $\sim$ sign implies that we equalize the left and the right hand sides up to terms $z^{n\geqslant0} w^m$ and $z^{n} w^{m\geqslant0}$.
The main ingredient of the commutation relations are the structure functions $\varphi^{(a,b)}(z)$ (that are also called bond factors \cite{Li:2023zub}):
\begin{equation}\label{bond_factors}
	\varphi^{(1,1)}(z) = \varphi^{(2,2)}(z) = \frac{z - \epsilon_1 - \epsilon_2}{z + \epsilon_1 + \epsilon_2}, \hspace{10mm} \varphi^{(1,2)}(z) = \varphi^{(2,1)}(z) = \frac{(z + \epsilon_1)(z + \epsilon_2)}{(z - \epsilon_1)(z - \epsilon_2)}\,.
\end{equation}

This is the place where the case $r=2$ deviates from generic $r$: in general the second factor also splits into two, appearing separately in $\varphi^{(a,a+1)}$ and $\varphi^{(a+1,a)}$:
$$
\varphi^{(a,a+1)}=\frac{z+\epsilon_1}{z-\epsilon_2},\quad \varphi^{(a+1,a)}=\frac{z+\epsilon_2}{z-\epsilon_1}\,.
$$
where the indices in the superscripts should be understood modulo $r$.
We will comment further on (de)construction of these functions from analogous expressions in the DIM algebra in Appendix \ref{sec:deco}.

One of the crucial properties of structure functions mentioned above is a reciprocity relation:
\begin{equation}
	\varphi^{(a,b)}(-z) = \frac{1}{\varphi^{(b,a)}(z)}\,.
\end{equation}
The above relations on generating functions can be represented as infinite families of relations between modes $\hat{e}^{(a)}_{n}, \hat{\psi}^{(a)}_{n}, \hat{f}^{(a)}_{n}$:
\begin{align}
	\label{rel psi}
	\Big[ \hat{\psi}^{(a)}_{n}, \hat{\psi}^{(b)}_{m} \Big] &= 0 \\
	\Big[ \hat{e}_{n}^{(a)}, \hat{f}_{m}^{(b)} \Big] &= \delta_{a,b} \, \hat{\psi}_{n+m}^{(a)}
\end{align}
Relations between raising $\hat{e}^{(a)}_{n}$ and lowering $\hat{f}^{(a)}_{n}$ generators of the same color $(a)$ read:
\begin{equation}
	\Big[ \hat{e}^{(a)}_{n+1}, \hat{e}^{(a)}_{m} \Big] - \Big[ \hat{e}^{(a)}_{n}, \hat{e}^{(a)}_{m+1} \Big] + (\epsilon_1 + \epsilon_2) \Big\{ \hat{e}^{(a)}_{n}, \hat{e}^{(a)}_{m},\Big\} = 0
\end{equation}
\begin{equation}
	\Big[ \hat{f}^{(a)}_{n+1}, \hat{f}^{(a)}_{m} \Big] - \Big[ \hat{f}^{(a)}_{n}, \hat{f}^{(a)}_{m+1} \Big] - (\epsilon_1 +\epsilon_2) \Big\{ \hat{f}^{(a)}_{n}, \hat{f}^{(a)}_{m},\Big\} = 0
\end{equation}
Relations of raising and lowering generators with $\hat{\psi}^{(a)}_{n}$ generators of the same color read:
\begin{equation}
	\Big[ \hat{\psi}^{(a)}_{n+1}, \hat{e}^{(a)}_{m} \Big] - \Big[ \hat{\psi}^{(a)}_{n}, \hat{e}^{(a)}_{m+1} \Big] + (\epsilon_1 + \epsilon_2) \Big\{ \hat{\psi}^{(a)}_{n}, \hat{e}^{(a)}_{m},\Big\} = 0
\end{equation}
\begin{equation}
	\Big[ \hat{\psi}^{(a)}_{n+1}, \hat{f}^{(a)}_{m} \Big] - \Big[ \hat{\psi}^{(a)}_{n}, \hat{f}^{(a)}_{m+1} \Big] - (\epsilon_1 +\epsilon_2) \Big\{ \hat{\psi}^{(a)}_{n}, \hat{f}^{(a)}_{m},\Big\} = 0
\end{equation}
The other class of relations involve modes of different colors $(a) \not= (b)$:
\begin{equation}
	\Big[ \hat{e}^{(a)}_{n+2}, \hat{e}^{(b)}_{m} \Big] - 2 \Big[ \hat{e}^{(a)}_{n+1}, \hat{e}^{(b)}_{m+1} \Big] + \Big[ \hat{e}^{(a)}_{n}, \hat{e}^{(b)}_{m+2} \Big] - (\epsilon_1 + \epsilon_2) \Bigg( \Big\{ \hat{e}^{(a)}_{n+1}, \hat{e}^{(b)}_{m}\Big\}  -  \Big\{ \hat{e}^{(a)}_{n}, \hat{e}^{(b)}_{m+1}\Big\} \Bigg) + \epsilon_1 \epsilon_2 \Big[\hat{e}^{(a)}_{n}, \hat{e}^{(b)}_{m} \Big] = 0
\end{equation}

\begin{equation}
	\Big[ \hat{f}^{(a)}_{n+2}, \hat{f}^{(b)}_{m} \Big] - 2 \Big[ \hat{f}^{(a)}_{n+1}, \hat{f}^{(b)}_{m+1} \Big] + \Big[ \hat{f}^{(a)}_{n}, \hat{f}^{(b)}_{m+2} \Big] +(\epsilon_1 + \epsilon_2) \Bigg( \Big\{ \hat{f}^{(a)}_{n+1}, \hat{f}^{(b)}_{m}\Big\}  -  \Big\{ \hat{f}^{(a)}_{n}, \hat{f}^{(b)}_{m+1}\Big\} \Bigg) + \epsilon_1 \epsilon_2 \Big[\hat{f}^{(a)}_{n}, \hat{f}^{(b)}_{m} \Big] = 0
\end{equation}

\begin{equation}
	\Big[ \hat{\psi}^{(a)}_{n+2}, \hat{e}^{(b)}_{m} \Big] - 2 \Big[ \hat{\psi}^{(a)}_{n+1}, \hat{e}^{(b)}_{m+1} \Big] + \Big[ \hat{\psi}^{(a)}_{n}, \hat{e}^{(b)}_{m+2} \Big] - (\epsilon_1 + \epsilon_2) \Bigg( \Big\{ \hat{\psi}^{(a)}_{n+1}, \hat{e}^{(b)}_{m}\Big\}  -  \Big\{ \hat{\psi}^{(a)}_{n}, \hat{e}^{(b)}_{m+1}\Big\} \Bigg) + \epsilon_1 \epsilon_2 \Big[\hat{\psi}^{(a)}_{n}, \hat{e}^{(b)}_{m} \Big] = 0
\end{equation}

\begin{equation}
	\Big[ \hat{\psi}^{(a)}_{n+2}, \hat{f}^{(b)}_{m} \Big] - 2 \Big[ \hat{\psi}^{(a)}_{n+1}, \hat{f}^{(b)}_{m+1} \Big] + \Big[ \hat{\psi}^{(a)}_{n}, \hat{f}^{(b)}_{m+2} \Big] +(\epsilon_1 + \epsilon_2) \Bigg( \Big\{ \hat{\psi}^{(a)}_{n+1}, \hat{f}^{(b)}_{m}\Big\}  -  \Big\{ \hat{\psi}^{(a)}_{n}, \hat{f}^{(b)}_{m+1}\Big\} \Bigg) + \epsilon_1 \epsilon_2 \Big[\hat{\psi}^{(a)}_{n}, \hat{f}^{(b)}_{m} \Big] = 0
\end{equation}

All these relations can be presented in the following form:
\begin{tcolorbox}
	\begin{equation}
		\label{short rel 1}
		\Big[ \hat{A}^{(a)}_{n+1}, \hat{B}^{(a)}_{m} \Big] - \Big[ \hat{A}^{(a)}_{n}, \hat{B}^{(a)}_{m+1} \Big] + \alpha_{(A,B)} \cdot (\epsilon_1 + \epsilon_2) \Big\{ \hat{A}^{(a)}_{n}, \hat{B}^{(a)}_{m},\Big\} = 0,
	\end{equation}
	\begin{align}
		\begin{aligned}
			\label{short rel 2}
			\Big[ \hat{A}^{(a)}_{n+2}, \hat{B}^{(b)}_{m} \Big] - 2 \Big[ \hat{A}^{(a)}_{n+1}, \hat{B}^{(b)}_{m+1} \Big] + \Big[ \hat{A}^{(a)}_{n}, \hat{B}^{(b)}_{m+2} \Big] +\epsilon_1 \epsilon_2 \Big[\hat{A}^{(a)}_{n}, \hat{B}^{(b)}_{m} \Big] - \\ - \alpha_{(A,B)} \cdot (\epsilon_1 + \epsilon_2) \Bigg( \Big\{ \hat{A}^{(a)}_{n+1}, \hat{B}^{(b)}_{m}\Big\}  -  \Big\{ \hat{A}^{(a)}_{n}, \hat{B}^{(b)}_{m+1}\Big\} \Bigg)  = 0,
		\end{aligned}
	\end{align}
\end{tcolorbox}
where $(A,B) \in \{(e,e), (\psi,e), (e,\psi), (f,f), (\psi,f), (f,\psi)\}$, and the the sign factor $\alpha_{(A,B)}$ takes values listed in Table \ref{tab:my_label 1}:
\begin{table}[h!]
	\centering
	\begin{tabular}{c|c|c|c|c|c|c}
		$(A,B)$ & $(e,e)$ &  $(\psi,e)$ & $(e,\psi)$ & $(f,f)$ & $(\psi,f)$ & $(f,\psi)$  \\
		\hline
		$\alpha_{(A,B)}$ & $1$ & $1$ & $1$ & $-1$ & $-1$ & $-1$ \\
	\end{tabular}
	\caption{Values of the $\alpha_{(A,B)}$}
	\label{tab:my_label 1}
\end{table}

There is a simple rule for $\alpha_{(A,B)}$: if any of the letters, $A$ or $B$, takes a value ``$f$'' then $\alpha_{(A,B)} = -1$.
These relations should be equipped with boundary conditions:
\begin{equation}
	\Big[ \hat{\psi}_{0}^{(a)}, \hat{e}_{n}^{(b)} \Big] = \mathcal{A}_{a,b} \, \hat{e}_{n}^{(b)}, \hspace{10mm} \Big[ \hat{\psi}_{0}^{(a)}, \hat{f}_{n}^{(b)} \Big] = - \mathcal{A}_{a,b} \, \hat{f}_{n}^{(b)}
\end{equation}
where $\mathcal{A}_{a,b}$ is a Cartan matrix of $\hat{\mathfrak{gl}}_{2}$:
\begin{equation}
	\mathcal{A}_{a,b} = \begin{pmatrix}
		\ 2 & -2 \\
		-2 & \ 2
	\end{pmatrix}
\end{equation}
The affine Yangian algebra contains a small set of operators that generate the other algebra in a similar way that simple roots do in Lie algebras. The set contains 6 operators:
\begin{equation}
	\label{small set}
	\hat{e}^{(1)}_{0}, \hspace{3mm} \hat{e}^{(2)}_{0}, \hspace{3mm} \hat{\psi}^{(1)}_{1}, \hspace{3mm} \hat{\psi}^{(2)}_{1}, \hspace{3mm} \hat{f}^{(1)}_{0}, \hspace{3mm} \hat{f}^{(2)}_{0}
\end{equation}
Higher modes $\hat{e}_{n}^{(a)},\hat{f}_{n}^{(a)}$ can be obtained with the help of Cartan generators $\hat{\psi}_{1}^{(a)},\hat{\psi}_{0}^{(a)}$ of the same color $(a)$:
\begin{equation}
	\label{raising mode a a}
	\hat{e}^{(a)}_{n+1} = \frac{1}{2} \, \Big[ \hat{\psi}_{1}^{(a)}, \hat{e}^{(a)}_{n} \Big] + \frac{(\epsilon_1 + \epsilon_2)}{2} \, \Big\{ \hat{\psi}_0^{(a)}, \hat{e}_{n}^{(a)} \Big\}, \hspace{10mm} \hat{f}^{(a)}_{n+1} = -\frac{1}{2} \, \Big[ \hat{\psi}_{1}^{(a)}, \hat{f}^{(a)}_{n} \Big] + \frac{(\epsilon_1 + \epsilon_2)}{2} \, \Big\{ \hat{\psi}_0^{(a)}, \hat{f}_{n}^{(a)} \Big\}
\end{equation}
or with Cartan generators of the other color $(b) \not= (a)$:
\begin{equation}
	\label{raising mode a b}
	\hat{e}^{(a)}_{n+1} = -\frac{1}{2} \, \Big[ \hat{\psi}_{1}^{(b)}, \hat{e}^{(a)}_{n} \Big] + \frac{(\epsilon_1 + \epsilon_2)}{2} \, \Big\{ \hat{\psi}_0^{(b)}, \hat{e}_{n}^{(a)} \Big\}, \hspace{10mm} \hat{f}^{(a)}_{n+1} = \frac{1}{2} \, \Big[ \hat{\psi}_{1}^{(b)}, \hat{f}^{(a)}_{n} \Big] + \frac{(\epsilon_1 + \epsilon_2)}{2} \, \Big\{ \hat{\psi}_0^{(b)}, \hat{f}_{n}^{(a)} \Big\}
\end{equation}
Using these formulas one can recursively obtain all the Yangian generators starting with the small set \eqref{small set} as the initial data. The zero mode Cartan generators $\hat{\psi}_0^{(a)}$ are not included in this small set because they are obtained from $\hat{e}^{(a)}_{0}, \hat{f}^{(a)}_{0}$ via \eqref{rel psi}. Strictly speaking, the operators $\hat{\psi}^{(1)}_{1},\hat{\psi}^{(2)}_{1}$ play a very similar role as one can see from \eqref{raising mode a a} and \eqref{raising mode a b}, therefore one of them is enough for the recursive procedure.

%%%%%%%%%%%%%%%%%%%%%%%%%%%%%%%%%%%%%%%%%%%%%%%%%%%%%%%%%%
%%%%%%%%%%%%%%%%%%%%%%%%%%%%%%%%%%%%%%%%%%%%%%%%%%%%%%%%%%
%%%%%%%%%%%%%%%%%%%%%%%%%%%%%%%%%%%%%%%%%%%%%%%%%%%%%%%%%%
%%%%%%%%%%%%%%%%%%%%%%%%%%%%%%%%%%%%%%%%%%%%%%%%%%%%%%%%%%

\section{Fock representation of $Y(\widehat{\mathfrak{gl}}_2)$} \label{sec:Fock reps}

\subsection{Crystal representations for $Y(\widehat{\fg\fl}_r)$} \label{sec:CryRep}
Young diagrams on the colored cell field (see Fig.~\ref{fig:QDglR}(c)) correspond naturally to 2d slices of 3d molten crystals enumerating vectors of Macmahon modules of $Y(\widehat{\mathfrak{gl}}_r)$.
Therefore we are able to implement the canonical crystal representation as an ansatz for the Fock module.

In this representation distinct vectors are in one-to-one correspondence with Young diagrams.
Let us denote as $\lambda^{\pm}$ sets of boxes that could be added/subtracted to/from diagram $\lambda$ so that the new diagram is again a Young diagram denoted as $\lambda\pm\Box$ respectively.
In what follows we will need to distinguish colors of the boxes.
We denote color $a$ of box $\Box$ as a letter inscribed in the box $\sqbox{$a$}$.
Also by denoting $\sqbox{$a$}\in \lambda$ we imply only boxes of color $a$ in set $\lambda$.

The Fock representation of $Y(\widehat{\mathfrak{gl}}_r)$ can be described by explicit matrices for $e_k^{(a)}$ and $f_k^{(a)}$:
\begin{equation}\label{cry_rep}
	\begin{aligned}
		e_k^{(a)}|\lambda\rangle&=\sum\lm_{\ssqbox{$a$}\in\lambda^+}{\bf E}_{\lambda,\lambda+\ssqbox{$a$}}\;\omega_{\ssqbox{$a$}}^k\,|\lambda+\sqbox{$a$}\rangle\,,\\
		f_k^{(a)}|\lambda\rangle&=\sum\lm_{\ssqbox{$a$}\in\lambda^-}{\bf F}_{\lambda,\lambda-\ssqbox{$a$}}\;\omega_{\ssqbox{$a$}}^k\,|\lambda-\sqbox{$a$}\rangle\,.
	\end{aligned}
\end{equation}
where $\omega_{\Box}$ is a box $\Box$ content defined by its coordinates (see Fig.~\ref{fig:QDglR}(c)):
\begin{equation}
	\omega_{\Box}=x_{\Box}\epsilon_1+y_{\Box}\epsilon_2\,.
\end{equation}

Mode-0 matrix coefficients can be calculated via products of diagram hooks (cf. \cite{FeiginTsymbaliuk}):
\begin{equation}\label{EF_cyc}
	\begin{aligned}
		{\bf E}_{\lambda,\lambda+\Box}&=\prod\lm_{\Box'\in\CH_{\lambda}(\Box)}\frac{\gamma_r\big({\rm arm}_{\lambda}(\Box'),{\rm leg}_{\lambda}(\Box')+1\big)}{\gamma_r\big({\rm arm}_{\lambda}(\Box')+1,{\rm leg}_{\lambda}(\Box')+1\big)}\prod\lm_{\Box'\in\CV_{\lambda}(\Box)}\frac{\gamma_r\big({\rm arm}_{\lambda}(\Box')+1,{\rm leg}_{\lambda}(\Box')\big)}{\gamma_r\big({\rm arm}_{\lambda}(\Box')+1,{\rm leg}_{\lambda}(\Box')+1\big)}\,,\\
		{\bf F}_{\lambda+\Box,\lambda}&=\prod\lm_{\Box'\in\CH_{\lambda}(\Box)}\frac{\gamma_r\big({\rm arm}_{\lambda}(\Box')+2,{\rm leg}_{\lambda}(\Box')\big)}{\gamma_r\big({\rm arm}_{\lambda}(\Box')+1,{\rm leg}_{\lambda}(\Box')\big)}\prod\lm_{\Box'\in\CV_{\lambda}(\Box)}\frac{\gamma_r\big({\rm arm}_{\lambda}(\Box'),{\rm leg}_{\lambda}(\Box')+2\big)}{\gamma_r\big({\rm arm}_{\lambda}(\Box'),{\rm leg}_{\lambda}(\Box')+1\big)}\,,
	\end{aligned}
\end{equation}
where vertical $\CV$ and horizontal $\CH$ strip sets for a new box as well as arm and leg lengths are defined in the canonical way (see \cite{FeiginTsymbaliuk} and Fig.~\ref{fig:vert_hor_arm_leg}), function $g_r(x,y)$ is defined as follows:
\begin{equation}\label{cyclic}
	\gamma_r(x,y):=\left\{\begin{array}{ll}
		-\epsilon_1 x+\epsilon_2 y, & \mbox{if }x+y=0\;{\rm mod}\;r;\\
		1,& \mbox{otherwise.}
	\end{array}\right.
\end{equation}

\begin{figure}
	\begin{center}
		\begin{tikzpicture}
			\begin{scope}
				\begin{scope}[scale=0.4]
					\foreach \x/\y/\z/\w in {0/-5/1/-5, 0/-4/0/-5, 0/-4/1/-4, 0/-3/0/-4, 0/-3/1/-3, 0/-2/0/-3, 0/-2/1/-2, 0/-1/0/-2, 0/-1/1/-1, 0/0/0/-1, 0/0/1/0, 1/-5/2/-5, 1/-4/1/-5, 1/-4/2/-4, 1/-3/1/-4, 1/-3/2/-3, 1/-2/1/-3, 1/-2/2/-2, 1/-1/1/-2, 1/-1/2/-1, 1/0/1/-1, 1/0/2/0, 2/-4/2/-5, 2/-4/3/-4, 2/-3/2/-4, 2/-3/3/-3, 2/-2/2/-3, 2/-2/3/-2, 2/-1/2/-2, 2/-1/3/-1, 2/0/2/-1, 2/0/3/0, 3/-3/3/-4, 3/-3/4/-3, 3/-2/3/-3, 3/-2/4/-2, 3/-1/3/-2, 3/-1/4/-1, 3/0/3/-1, 3/0/4/0, 4/-3/5/-3, 4/-2/4/-3, 4/-2/5/-2, 4/-1/4/-2, 4/-1/5/-1, 4/0/4/-1, 4/0/5/0, 5/-3/6/-3, 5/-2/5/-3, 5/-2/6/-2, 5/-1/5/-2, 5/-1/6/-1, 5/0/5/-1, 5/0/6/0, 6/-2/6/-3, 6/-2/7/-2, 6/-1/6/-2, 6/-1/7/-1, 6/0/6/-1, 6/0/7/0, 7/-1/7/-2, 7/-1/8/-1, 7/0/7/-1, 7/0/8/0, 8/-1/9/-1, 8/0/8/-1, 8/0/9/0, 9/0/9/-1}
					{
						\draw (\x,\y) -- (\z,\w);
					}
					\draw[fill=palette7] (3,-3) -- (4,-3) -- (4,-4) -- (3,-4) -- cycle;
					\foreach \i in {0,1,2}
					{
						\draw[fill=palette2] (\i,-3) -- (\i+1,-3) -- (\i+1,-4) -- (\i,-4) -- cycle;
					}
					\node[left,palette2] at (-2,-3.5) {$\CH_{\lambda}({\color{palette7}\blacksquare})$};
					\draw[-stealth] (-1.8,-3.5) -- (-0.2,-3.5);
					\foreach \i in {0,1,2}
					{
						\draw[fill=palette4] (3,-\i) -- (3,-\i-1) -- (4,-\i-1) -- (4,-\i) -- cycle;
					}
					\node[right,palette4] at (6,-3.5) {$\CV_{\lambda}({\color{palette7}\blacksquare})$};
					\draw[-stealth] (5.8,-3.5) to[out=180,in=0] (4.2,-1.5);
				\end{scope}
			\end{scope}
			%%%%%%%%%%%%%%%%%%%%
			\begin{scope}[shift={(7,0)}]
				\begin{scope}[scale=0.4]
					\foreach \x/\y/\z/\w in {0/-5/1/-5, 0/-4/0/-5, 0/-4/1/-4, 0/-3/0/-4, 0/-3/1/-3, 0/-2/0/-3, 0/-2/1/-2, 0/-1/0/-2, 0/-1/1/-1, 0/0/0/-1, 0/0/1/0, 1/-5/2/-5, 1/-4/1/-5, 1/-4/2/-4, 1/-3/1/-4, 1/-3/2/-3, 1/-2/1/-3, 1/-2/2/-2, 1/-1/1/-2, 1/-1/2/-1, 1/0/1/-1, 1/0/2/0, 2/-4/2/-5, 2/-4/3/-4, 2/-3/2/-4, 2/-3/3/-3, 2/-2/2/-3, 2/-2/3/-2, 2/-1/2/-2, 2/-1/3/-1, 2/0/2/-1, 2/0/3/0, 3/-3/3/-4, 3/-3/4/-3, 3/-2/3/-3, 3/-2/4/-2, 3/-1/3/-2, 3/-1/4/-1, 3/0/3/-1, 3/0/4/0, 4/-3/5/-3, 4/-2/4/-3, 4/-2/5/-2, 4/-1/4/-2, 4/-1/5/-1, 4/0/4/-1, 4/0/5/0, 5/-3/6/-3, 5/-2/5/-3, 5/-2/6/-2, 5/-1/5/-2, 5/-1/6/-1, 5/0/5/-1, 5/0/6/0, 6/-2/6/-3, 6/-2/7/-2, 6/-1/6/-2, 6/-1/7/-1, 6/0/6/-1, 6/0/7/0, 7/-1/7/-2, 7/-1/8/-1, 7/0/7/-1, 7/0/8/0, 8/-1/9/-1, 8/0/8/-1, 8/0/9/0, 9/0/9/-1}
					{
						\draw (\x,\y) -- (\z,\w);
					}
					\draw[fill=palette6] (1,-1) -- (2,-1) -- (2,-2) -- (1,-2) -- cycle;
					\draw[fill=palette3] (1,-2) -- (2,-2) -- (2,-5) -- (1,-5) -- cycle;
					\node[rotate=90,white] at (1.5,-3.5) {\small leg};
					\draw[fill=palette1] (2,-1) -- (7,-1) -- (7,-2) -- (2,-2) -- cycle;
					\node[white] at (4.5,-1.5) {\small arm};
				\end{scope}
			\end{scope}
		\end{tikzpicture}
	\end{center}
	\caption{Vertical, horizontal sets, leg and arm functions on Young diagrams.}\label{fig:vert_hor_arm_leg}
\end{figure}

A natural norm on the vectors of the Fock representation reads:
\begin{equation}\label{Fock_norm}
	\langle\lambda|\mu\rangle=\delta_{\lambda,\mu}\times{\bf N}_{\lambda}\,.
\end{equation}
where
\begin{equation}
	{\bf N}_{\lambda}=\prod\lm_{\Box\in\lambda}\gamma_r\big({\rm arm}_{\lambda}(\Box)+1,{\rm leg}_{\lambda}(\Box)\big)\,\gamma_r\big({\rm arm}_{\lambda}(\Box),{\rm leg}_{\lambda}(\Box)+1\big)\,.
\end{equation}
With respect to this norm generators of opposite Borel parity are conjugated:
\begin{equation}
	\left(f_k^{(a)}\right)^*=e_k^{(a)}\,.
\end{equation}

Furthermore we observe for $r=2$:
\begingroup
\renewcommand*{\arraystretch}{1.8}
\begin{tcolorbox}
	\begin{equation}\label{triality}
		\begin{array}{ccccc}
			\mbox{\small\color{burgundy} Algebra} && \mbox{\small\color{burgundy} Polynomials} && \mbox{\small\color{burgundy} Quiver}  \\
			\hline
			{\bf N}_{\lambda}&=&\left\langle U_{\lambda}^{(2)},U_{\lambda}^{(2)}\right\rangle_2 &=&{\rm Eul}_{\lambda}^{\widehat{\fg\fl}_2}\,,\\ [5pt]
			%%%%%%%%%%%%%%%%%%%%%%%%%%%%%%%%%%%%%%%%%%%%%%%
			{\bf E}_{\lambda,\lambda+\ssqbox{$a$}}&=&\dfrac{\left\langle U_{\lambda+\ssqbox{$a$}}^{(2)},p_1\cdot U_{\lambda}^{(2)}\right\rangle_2}{\left\langle U_{\lambda+\ssqbox{$a$}}^{(2)},U_{\lambda+\ssqbox{$a$}}^{(2)}\right\rangle_2}&=&\dfrac{{\rm Eul}_{\lambda}^{\widehat{\fg\fl}_2}}{{\rm Eul}_{\lambda,\lambda+\ssqbox{$a$}}^{\widehat{\fg\fl}_2}}\,,\\[15pt]
			%%%%%%%%%%%%%%%%%%%%%%%%%%%%%%%%%%%%%%%%%%%%%%%
			{\bf F}_{\lambda,\lambda-\ssqbox{$a$}}&=&\dfrac{\left\langle U^{(2)}_{\lambda-\ssqbox{$a$}},\frac{\p}{\p p_1}\cdot U^{(2)}_{\lambda}\right\rangle_2}{\left\langle U^{(2)}_{\lambda-\ssqbox{$a$}},U^{(2)}_{\lambda-\ssqbox{$a$}}\right\rangle_2}&=&\dfrac{{\rm Eul}_{\lambda}^{\widehat{\fg\fl}_2}}{{\rm Eul}_{\lambda-\ssqbox{$a$},\lambda}^{\widehat{\fg\fl}_2}}\,.
		\end{array}
	\end{equation}
	%{\bf\color{burgundy} up to signs!!!} and {\bf\color{burgundy} satisfy} \eqref{relations}.
\end{tcolorbox}
\endgroup

In these relations the set of quantities denoted ${\rm Eul}_{\lambda}^{\mathfrak{Q}}$ and ${\rm Eul}_{\lambda,\lambda+\Box}^{\mathfrak{Q}}$ are equivariant Euler classes of the tangent bundles to the corresponding fixed points labeled by partitions $\lambda$ on a quiver variety labeled by $\mathfrak{Q}$ (for the identification of a Dynkin diagram for $\widehat{\fg\fl}_r$ and a quiver variety see Fig.~\ref{fig:QDglR}).
A single Young diagram $\lambda$ labels a fixed point on the quiver representation -- a classical vacuum, whereas a pair of diagrams $\lambda,\mu$ labels a fixed point on an incidence locus of homomorphisms between pairs of representations.
The structure of the matrix coefficients and the noramlization \eqref{triality} for the BPS algebra of D-branes on a toric Calabi-Yau 3-fold is standard and discussed in the literature extensively \cite{Galakhov:2020vyb,Galakhov:2021vbo,Rapcak:2020ueh}.
Therefore we will not comment on this calculation here rather referring the reader to the notations and the setup of \cite{Galakhov:2023mak} and an explicit example of such a calculation in Appendix A.3 therein.

Relations \eqref{triality} establish a version of a \emph{triality} between different constructions of the affine Yangian algebras.
Eventually we expect that the representations of $r$-Uglov polynomials are equivalent -- have identical operator matrix elements -- rather than simply homomorphic to BPS wave function representations of algebras $Y(\widehat{\fg\fl}_r)$, and both form  well-established Fock modules of $Y(\widehat{\fg\fl}_r)$, where the matrix entry for the $(\lambda,\lambda+\Box)$-index is given by an explicit expression.

%%%%%%%%%%%%%%%%%%%%%%%%%%%%%%%%%%%%%%%%%%%%%%%%%%
%%%%%%%%%%%%%%%%%%%%%%%%%%%%%%%%%%%%%%%%%%%%%%%%%%
%%%%%%%%%%%%%%%%%%%%%%%%%%%%%%%%%%%%%%%%%%%%%%%%%%
%%%%%%%%%%%%%%%%%%%%%%%%%%%%%%%%%%%%%%%%%%%%%%%%%%

\subsection{Time operators and Yangian generators}\label{sec:time_op}

The approach to the Fock representations of $Y(\widehat{\fg\fl}_r)$ via the Young diagram bases does not seem to produce any a priori time variables $p_k$.
Therefore we try to mimic them by operators $\hat \xi_k$ that represent a trivial operator of multiplication by $p_k\cdot$ in terms of the Uglov polynomials.
To summarize we are going to search for operators $\hat \xi_k$ with the following properties:
\begin{tcolorbox}
	\begin{equation}
		\begin{aligned}
			\mbox{a)} & \quad \quad \left[\hat\xi_i,\hat\xi_j\right]=0,\quad \forall\, i,j\,,\\
			\mbox{b)} & \quad \quad \hat\xi_k\cdot U^{(r)}_{\lambda}(p_1,p_2,\ldots)\Big|_{p_m\to\hat\xi_m}=\left(p_k\,U^{(r)}_{\lambda}(p_1,p_2,\ldots)\right)\Big|_{p_m\to\hat\xi_m}\,,\\
			\mbox{c)} & \quad \quad U^{(r)}_{\lambda}(p_1,p_2,\ldots)\Big|_{p_m\to\hat\xi_m}\,|\varnothing\rangle\,=\,|\lambda\rangle\,.
		\end{aligned}
	\end{equation}
\end{tcolorbox}
In this subsection we also restrict ourselves to the case $r=2$ as the simplest instance of the time operator inexpressibility as we will see in what follows.

We represent the time operators as consequent processes of adding boxes. 
Each elementary adding event is accompanied by an amplitude from \eqref{cry_rep}.
Hence we calculate:
\begin{equation}\label{xi-times}
	\begin{aligned}
		\hat\xi_1|\lambda\rangle&=\sum\lm_{\ssqbox{$a$}}{\bf E}_{\lambda,\lambda+\ssqbox{$a$}}|\lambda+\sqbox{$a$}\rangle\,,\\
		\hat\xi_2|\lambda\rangle&=\epsilon_1\sum\lm_{\ssqbox{$a$},\ssqbox{$b$}}\frac{{\bf E}_{\lambda,\lambda+\ssqbox{$a$}}{\bf E}_{\lambda+\ssqbox{$a$},\lambda+\ssqbox{$a$}+\ssqbox{$b$}}}{\omega_{\ssqbox{$b$}}-\omega_{\ssqbox{$a$}}}(1-\delta_{a,b})|\lambda+\sqbox{$a$}+\sqbox{$b$}\rangle\,,\\
		\hat\xi_3|\lambda\rangle&=\sum\lm_{\ssqbox{$a$},\ssqbox{$b$},\ssqbox{$c$}}\Xi_{a,b,c}{\bf E}_{\lambda,\lambda+\ssqbox{$a$}}{\bf E}_{\lambda+\ssqbox{$a$},\lambda+\ssqbox{$a$}+\ssqbox{$b$}}{\bf E}_{\lambda+\ssqbox{$a$}+\ssqbox{$b$},\lambda+\ssqbox{$a$}+\ssqbox{$b$}+\ssqbox{$c$}}|\lambda+\sqbox{$a$}+\sqbox{$b$}+\sqbox{$c$}\rangle\,,
	\end{aligned}
\end{equation}
where the color tensor reads:
\begin{equation}
	\Xi_{1,2,1}=\Xi_{2,1,2}=1,\quad \Xi_{1,2,2}=\Xi_{2,2,1}=\Xi_{2,1,1}=\Xi_{1,1,2}=-\frac{1}{2},\quad \Xi_{1,1,1}=\Xi_{2,2,2}=0\,.
\end{equation}
In this calculation we observe the following phenomena.
Odd time operators are given by regular polynomial expressions in $e_0^{(a)}$:
\begin{equation}
	\hat \xi_{2k-1}=\frac{1}{2^k}\bigg(\underbrace{\left[e_0^{(1)},\left[e_0^{(2)},\ldots\left[e_0^{(2)},e_0^{(1)}\right]\ldots\right]\right]}_{2k-1{\rm \; times}}+\underbrace{\left[e_0^{(2)},\left[e_0^{(1)},\ldots\left[e_0^{(1)},e_0^{(2)}\right]\ldots\right]\right]}_{2k-1{\rm \; times}}\bigg),\quad k=1,2,\ldots
\end{equation}

Whereas for even times there are no regular expressions due to a pole.
We could use other operators of $Y(\widehat{\fg\fl}_2)$ to regularize expressions for $\hat\xi_2$.
Eventually we arrive to the following regular relations:
\begin{equation}\label{even_reg}
	\frac{1}{\epsilon_1+\epsilon_2}\left[\psi_1^{(2)},\hat\xi_{2k}\right]+4k(\epsilon_1+\epsilon_2)\hat\xi_{2k}=\epsilon_1\underbrace{\left[e_0^{(2)},\left[e_0^{(1)},\ldots\left[e_0^{(2)},e_0^{(1)}\right]\ldots\right]\right]}_{2k{\rm \; times}},\quad k=1,2,\ldots
\end{equation}

In particular:
\begin{equation}
	\begin{aligned}
		\hat \xi_1&=e_0^{(1)}+e_0^{(2)}\,,\\
		\frac{1}{\epsilon_1+\epsilon_2}\left[\psi_1^{(2)},\hat\xi_2\right]+4(\epsilon_1+\epsilon_2)\hat\xi_2&=\epsilon_1\left[e_0^{(2)},e_0^{(1)}\right]\,,\\
		\hat \xi_3&=\frac{1}{2}\left(\left[e_0^{(1)},\left[e_0^{(2)},e_0^{(1)}\right]\right]+\left[e_0^{(2)},\left[e_0^{(1)},e_0^{(2)}\right]\right]\right)\,,\\
		\frac{1}{\epsilon_1+\epsilon_2}\left[\psi_1^{(2)},\hat\xi_4\right]+8(\epsilon_1+\epsilon_2)\hat\xi_4&=\epsilon_1\left[e_0^{(2)},\left[e_0^{(1)},\left[e_0^{(2)},e_0^{(1)}\right]\right]\right]\,,\\
		\hat \xi_5&=\frac{1}{4}\left(\left[e_0^{(1)},\left[e_0^{(2)},\left[e_0^{(1)},\left[e_0^{(2)},e_0^{(1)}\right]\right]\right]\right]+\left[e_0^{(2)},\left[e_0^{(1)},\left[e_0^{(2)},\left[e_0^{(1)},e_0^{(2)}\right]\right]\right]\right]\right)\,,\\
	\end{aligned}
\end{equation}

As we have seen in \eqref{raising mode a a} and \eqref{raising mode a b} operator $\psi_1^{(a)}$ raises necessarily the modes of operators $e_k^{(a)}$ and $f_k^{(a)}$.
Therefore the r.h.s. of \eqref{even_reg} with only operators in mode 0 belongs to a cokernel of the linear operator in the l.h.s.
Therefore relations \eqref{even_reg} have no solutions for operators $\hat\xi_{2k}$ in terms of $Y(\widehat{\fg\fl}_2)$.

\bigskip

The inexpressibility of even times in terms of $Y(\widehat{\fg\fl}_2)$ generators
may not be yet familiar in Yangian research
(though it is well known in general representation theory,
see an example at the end of this subsection).
%and might seem somewhat absurd at the first sight.
In this paragraph we explain, why the problem can disappear in the limit
%Moreover, further we will consider a limit
$\beta=1$ ($\epsilon_1=1$, $\epsilon_2=-1$).
In this limit Uglov polynomials $U_{\lambda}^{(r)}$ for all $r$ become simply the Schur polynomials for the corresponding diagrams $\lambda$.
This seeming contradiction is resolved by \emph{new} relations (sometimes even mixing different algebras $Y(\widehat{\fg\fl}_r)$) that cease to be present at $\beta\neq 1$.
The basic simplification occurring at $\beta=1$ is for the matrix coefficients:
\begin{equation}
	{\bf E}_{\lambda,\lambda+\Box}^{(\beta=1)}=1\,.
\end{equation}
The pole term in $\hat \xi_2$ in \eqref{xi-times} is anti-symmetric with respect to the permutation of the box adding order.
Therefore the only non-trivial terms appearing in this sum correspond to situations when the box adding order is not free: when we add to partition $\lambda$ a 2-box cluster \!\!\raisebox{-0.3ex}{$
	\begin{array}{c}
		\begin{tikzpicture}[scale=0.25]
			\foreach \x/\y/\z/\w in {0/-2/1/-2, 0/-1/0/-2, 0/-1/1/-1, 0/0/0/-1, 0/0/1/0, 1/-1/1/-2, 1/0/1/-1}
			{
				\draw (\x,\y) -- (\z,\w);
			}
			\node at (0.5,-0.5) {$\scriptstyle a$};
			\node at (0.5,-1.5) {$\scriptstyle b$};
		\end{tikzpicture}
	\end{array}$}\!\! or \!\!\raisebox{-0.3ex}{$\begin{array}{c}
	\begin{tikzpicture}[scale=0.25]
		\foreach \x/\y/\z/\w in {0/-1/1/-1, 0/0/0/-1, 0/0/1/0, 1/-1/2/-1, 1/0/1/-1, 1/0/2/0, 2/0/2/-1}
		{
			\draw (\x,\y) -- (\z,\w);
		}
		\node at (0.5,-0.5) {$\scriptstyle a$};
		\node at (1.5,-0.5) {$\scriptstyle b$};
	\end{tikzpicture}
	\end{array}$}\!\!, $a\neq b$.
In this situation we can forget about colors due to chess-board coloring of partitions in $Y(\fg\fl_2)$, and the difference of box contents is a mere sign:
\begin{equation}
	(\omega_{\ssqbox{$a$}}-\omega_{\ssqbox{$b$}})\Big[\!\!\raisebox{-0.3ex}{$
		\begin{array}{c}
		\begin{tikzpicture}[scale=0.25]
			\foreach \x/\y/\z/\w in {0/-2/1/-2, 0/-1/0/-2, 0/-1/1/-1, 0/0/0/-1, 0/0/1/0, 1/-1/1/-2, 1/0/1/-1}
			{
				\draw (\x,\y) -- (\z,\w);
			}
			\node at (0.5,-0.5) {$\scriptstyle a$};
			\node at (0.5,-1.5) {$\scriptstyle b$};
		\end{tikzpicture}
		\end{array}$}\!\!\Big]=\pm 1,\quad (\omega_{\ssqbox{$a$}}-\omega_{\ssqbox{$b$}})\big[
		\!\!\raisebox{-0.3ex}{$\begin{array}{c}
		\begin{tikzpicture}[scale=0.25]
			\foreach \x/\y/\z/\w in {0/-1/1/-1, 0/0/0/-1, 0/0/1/0, 1/-1/2/-1, 1/0/1/-1, 1/0/2/0, 2/0/2/-1}
			{
				\draw (\x,\y) -- (\z,\w);
			}
			\node at (0.5,-0.5) {$\scriptstyle a$};
			\node at (1.5,-0.5) {$\scriptstyle b$};
		\end{tikzpicture}
	\end{array}$}\!\!\big]=\pm 1\,.
\end{equation}
Having noticed this we arrive to the following relation:
\begin{equation}\label{beta=1_xi_2}
	\begin{aligned}
		&\sum\lm_{\ssqbox{$a$},\ssqbox{$b$}}\frac{{\bf E}_{\lambda,\lambda+\ssqbox{$a$}}^{(\beta=1)}{\bf E}_{\lambda+\ssqbox{$a$},\lambda+\ssqbox{$a$}+\ssqbox{$b$}}^{(\beta=1)}}{\omega_{\ssqbox{$a$}}-\omega_{\ssqbox{$b$}}}|\lambda\rangle=\sum\lm_{\ssqbox{$a$},\ssqbox{$b$}}\left(\omega_{\ssqbox{$a$}}-\omega_{\ssqbox{$b$}}\right){\bf E}_{\lambda,\lambda+\ssqbox{$a$}}^{(\beta=1)}{\bf E}_{\lambda+\ssqbox{$a$},\lambda+\ssqbox{$a$}+\ssqbox{$b$}}^{(\beta=1)}|\lambda\rangle=\\
		&=\sum\lm_{\Box,\Box'}\left(\omega_{\Box}-\omega_{\Box'}\right){\bf E}_{\lambda,\lambda+\Box}^{(\beta=1)}{\bf E}_{\lambda+\Box,\lambda+\Box+\Box'}^{(\beta=1)}|\lambda+\Box+\Box'\rangle=\left[e_0,e_1\right]|\lambda\rangle\,,
	\end{aligned}
\end{equation}
where $e_k$ are colorless generators of $Y(\widehat{\fg\fl}_1)$ (see \eqref{cry_rep}) (containing sums over all new boxes regardless their actual color).

Another difference from the $\beta=1$ case is that the standard $Y(\widehat{\fg\fl}_1)$ is not a subalgebra
of other ``less-Abelian'' $Y(\widehat{\fg\fl}_r)$ when $\beta\neq 1$.
In particular, the operator $ \sum ap_{a+1} \frac{\p}{\p p_a} \notin  Y(\widehat{\fg\fl}_r)$ for $r>1$ and $\beta\neq 1$,
thus it can not help to generate higher time-variables from canonically defined $p_1=\sum_{i=1}^r e_0^{(i)}$,
as it does in the Jack representation of $Y(\widehat{\fg\fl}_1)$.

\bigskip

Thus we see that associating the time variables with a commuting subalgebra in $Y(\widehat{\fg\fl}_r)$
for $r>1$ might be a problem.
Still this does not contradict a possibility to represent the algebra through differential operators in times.
Just this representation can be a little too big, and Yangian itself is a kind of a factor.
This can be similar to the story with bosonisation of  Kac-Moody algebras \cite{GMMOS},
where the algebra commutes with the screening charges, which are non-trivial functions of the free fields and times.
In the next Section we describe the simplest such bosonisation of $Y(\widehat{\fg\fl}_2)$.

%%%%%%%%%%%%%%%%%%%%%%%%%%%%%%%%%%%%%%%%%%%%%%%%%%%%%%
%%%%%%%%%%%%%%%%%%%%%%%%%%%%%%%%%%%%%%%%%%%%%%%%%%%%%%
%%%%%%%%%%%%%%%%%%%%%%%%%%%%%%%%%%%%%%%%%%%%%%%%%%%%%%

\section{Time-variable representation} \label{sec: time var reps}

\subsection{Special case $\epsilon_1 + \epsilon_2 = 0$, i.e. $\beta = 1$}
In this representation the states of the Fock representation are in one-to-one correspondence with the Schur polynomials:
\begin{equation}
	\ket{ \lambda } = S_{\lambda} \{ p \}
\end{equation}
where the Schur polynomials in variables $p_a, a \in \mathbb{N}$ are defined in the following way:
\begin{equation}
	\exp\left( \sum^{\infty}_{k=0} \frac{p_k}{k} z^k \right) = \sum^{\infty}_{n=0} z^n \cdot S_{[n]}, \hspace{15mm} S_{\lambda} = \det_{1 \leqslant i,j \leqslant l(\lambda)} \left( S_{[\lambda_i + j - i]} \right)
\end{equation}
where $[n]$ is a one row Young diagram. We provide explicit formulas for the small set \eqref{small set}, and the other generators of the Yangian can be obtained with the help of the small set. The first are the raising operators:
\begin{equation}
	\hat{e}_{0}^{(1)} = p_1 + \sum_{n=1}^{\infty}(-)^n \sum_{i=1}^{n+1} \sum_{j=1}^{n} S_{(i|n+1-i)} \hat{S}_{(j|n-j)} = p_1 - p_1^2 \frac{\partial}{\partial p_1} + \frac{(2 p_1^3 + p_3)}{3} \frac{\partial^2}{\partial p_1^2}  +\ldots
\end{equation}
\begin{equation}
	\hat{e}_{0}^{(2)} = - \sum^{\infty}_{n=1}(-)^n \sum_{i=1}^{n+1} \sum_{j=1}^{n}  S_{(i|n+1-i)} \hat{S}_{(j|n-j)} =  p_1^2 \frac{\partial}{\partial p_1} - \frac{(2 p_1^3 + p_3)}{3} \frac{\partial^2}{\partial p_1^2} +\ldots
\end{equation}
They increase the number of boxes in the colored Young diagram by one:
\begin{equation}
    \hat{e}_{0}^{(a)} S_{\lambda} = \sum_{\ssqbox{$a$} \in \lambda^{+}} S_{\lambda + \ssqbox{$a$}}
\end{equation}
Let us note that the coefficients in the r.h.s. in front of the Schur polynomials are unit and it is a feature of the $\beta = 1$ representation. Lowering operators
\begin{equation}
	\hat{f}_{0}^{(1)} = \frac{\partial}{\partial p_1} + \sum_{n=1}^{\infty} (-)^n \sum_{i=1}^{n} \sum_{j=1}^{n+1} S_{(i|n-i)} \hat{S}_{(j|n+1-j)} = \frac{\partial}{\partial p_1} - p_1 \frac{\partial^2}{\partial p_1^2} + \frac{p_1^2}{3} \left( 2\frac{\partial^3}{\partial p_1^3}+3\frac{\partial}{\partial p_3}\right) + \ldots
\end{equation}
\begin{equation}
	\hat{f}_{0}^{(2)} = - \sum_{n=1} (-)^n \sum_{i=1}^{n} \sum_{j=1}^{n+1} S_{(i|n-i)} \hat{S}_{(j|n+1-j)} = p_1 \frac{\partial^2}{\partial p_1^2} - \frac{p_1^2}{3} \left( 2\frac{\partial^3}{\partial p_1^3}+3\frac{\partial}{\partial p_3}\right) + \ldots
\end{equation}
remove boxes from the colored Young diagrams:
\begin{equation}
    \hat{f}_0^{(a)} S_{\lambda} = \sum_{\ssqbox{$a$} \in \lambda^{-}} S_{\lambda - \ssqbox{$a$}}
\end{equation}
The sum over colors gives the corresponding operators of the $Y(\hat{\mathfrak{gl}}_1)$:
\begin{equation}
    \hat{e}^{(1)}_{0} + \hat{e}^{(2)}_{0} = p_1, \hspace{15mm} \hat{f}^{(1)}_{0} + \hat{f}^{(2)}_{0} = \frac{\partial}{\partial p_1}
\end{equation}
The commutator of the zero modes of the raising and lowering operators gives a grading operator:
\begin{equation}
    \Big[ \hat{e}_0^{(1)}, \hat{f}_0^{(1)} \Big] = \hat{\psi}_{0}^{(1)} = -1 - 2 \sum_{n=1}^{\infty} (-)^n \sum_{i=1}^{n} \sum_{j=1}^{n} S_{(i|n-i)} \hat{S}_{(j|n-j)}
\end{equation}
\begin{equation}
    \Big[ \hat{e}_0^{(2)}, \hat{f}_0^{(2)} \Big] = \hat{\psi}_{0}^{(2)} = 2 \sum_{n=1}^{\infty} (-)^n \sum_{i=1}^{n} \sum_{j=1}^{n} S_{(i|n-i)} \hat{S}_{(j|n-j)}
\end{equation}
And the sum of these operators is trivial:
\begin{equation}
    \hat{\psi}_{0}^{(1)} + \hat{\psi}_{0}^{(2)} = -1
\end{equation}
Simpler Yangian algebra $Y(\hat{\mathfrak{gl}}_1)$ contains grading operator $\hat{\psi}_2$ with eigenvalues $n_{\lambda} = \sum_{\Box \in \lambda} 1$. In the $Y(\hat{\mathfrak{gl}}_2)$ case the eigenvalues respect colors of boxes so that they contribute with opposite signs:
\begin{equation}
    \hat{\psi}_{0}^{(1)} S_{\lambda} = \left( -1 + 2 n_{\lambda}^{(1)} - 2 n_{\lambda}^{(2)} \right) S_{\lambda}, \hspace{10mm} \hat{\psi}_{0}^{(2)} S_{\lambda} = \left( 2 n_{\lambda}^{(2)} - 2 n_{\lambda}^{(1)} \right) S_{\lambda},
\end{equation}
Note that the above eigenvalues depend only on the differences of number of boxes of different colors:
\begin{equation}
    n_{\lambda}^{(a)} = \sum_{\ssqbox{$a$} \in \lambda} 1
\end{equation}
The main ingredient of the representation are the operators $\hat{\psi}_1^{(1)}$ and $\hat{\psi}_1^{(2)}$. They represent a counterpart of the cut-and-join operator $\hat{\psi}_3$ from $Y(\hat{\mathfrak{gl}}_1)$:
\begin{align}
	\begin{aligned}
		\hat{\psi}_{1}^{(1)} &= - 2 \sum^{\infty}_{n=1} (-)^n \sum_{i=1}^{n}  \sum_{j=1}^{n} (n+1-i-j) \, S_{(i|n-i)} \hat{S}_{(j|n-j)} = \\&=  - p_2 \frac{\partial^2}{\partial p_1^2} - 2 p_1^2 \frac{\partial}{\partial p_2} + \frac{4(2 p_1^3 + p_3)}{9} \frac{\partial^2}{\partial p_1 \partial p_2} + 2 p_1 p_2 \frac{\partial}{\partial p_3} + \frac{4p_1 p_2}{3} \frac{\partial^3}{\partial p_1^3} + \ldots
	\end{aligned}
\end{align}
The second operator $\hat{\psi}_{1}^{(2)} = - \hat{\psi}_{1}^{(1)}$ and this formula is specific for $\beta = 1$ case. The eigenvalues are the sum of contents $ x_{\Box}-y_{\Box}$ over the diagram with signs that respect colors:
\begin{equation}
    \hat{\psi}_{1}^{(1)} S_{\lambda} = 2 \left(\kappa^{(1)}_{\lambda} -\kappa^{(2)}_{\lambda} \right) S_{\lambda}, \hspace{10mm} \hat{\psi}_{1}^{(2)} S_{\lambda} = 2 \left(\kappa^{(2)}_{\lambda} -\kappa^{(1)}_{\lambda} \right) S_{\lambda}
\end{equation}
where we introduce quantity $\kappa_{\lambda}^{(a)}$:
\begin{equation}
    \kappa_{\lambda}^{(a)} = \sum_{\ssqbox{$a$} \in \lambda} x_{\ssqbox{$a$}} - y_{\ssqbox{$a$}}
\end{equation}
In our notations the first box of the Young diagram has coordinates $(x,y) = (0,0)$ and $x$ is a horizontal coordinate (see Fig.\ref{fig:QDglR}). $\hat{\psi}^{(1)}_{1}$ operator is crucial in our construction due to the fact that it allows us to increase the mode number of raising/lowering operators:
\begin{equation}
    \hat{e}^{(1)}_{n+1} = \frac{1}{2} \Big[  \hat{\psi}^{(1)}_{1},  \hat{e}^{(1)}_{n}\Big], \hspace{10mm} \hat{e}^{(2)}_{n+1} = -\frac{1}{2} \Big[  \hat{\psi}^{(1)}_{1},  \hat{e}^{(2)}_{n}\Big]
\end{equation}
\begin{equation}
    \hat{f}^{(1)}_{n+1} = -\frac{1}{2} \Big[  \hat{\psi}^{(1)}_{1},  \hat{f}^{(1)}_{n}\Big], \hspace{10mm} \hat{f}^{(2)}_{n+1} = -\frac{1}{2} \Big[  \hat{\psi}^{(1)}_{1},  \hat{f}^{(2)}_{n}\Big]
\end{equation}
The explicit form of the first mode raising operators reads:
\begin{equation}\label{e_1_1_beta=1}
    \hat{e}_{1}^{(1)} = \sum_{n=1}^{\infty}  \sum_{i=1}^{n+1} \sum_{j=1}^{n} \ \Big[(-)^n(n + 3/2 - i - j) + 1/2 (-)^{i+j} \Big] \ S_{(i|n+1-i)} \hat{S}_{(j|n-j)}
\end{equation}
\begin{equation}\label{e_1_2_beta=1}
    \hat{e}_{1}^{(2)} = \sum_{n=1}^{\infty}  \sum_{i=1}^{n+1} \sum_{j=1}^{n} \ \Big[-(-)^n(n + 3/2 - i - j) + 1/2 (-)^{i+j} \Big] \ S_{(i|n+1-i)} \hat{S}_{(j|n-j)}
\end{equation}

The lowering operators have a similar structure:
\begin{equation}
    \hat{f}_{1}^{(1)} = \sum_{n=1}^{\infty}  \sum_{i=1}^{n+1} \sum_{j=1}^{n} \ \Big[(-)^n(n + 3/2 - i - j) + 1/2 (-)^{i+j} \Big] \ S_{(i|n-i)} \hat{S}_{(j|n+1-j)}
\end{equation}
\begin{equation}
    \hat{f}_{1}^{(2)} = \sum_{n=1}^{\infty}  \sum_{i=1}^{n+1} \sum_{j=1}^{n} \ \Big[-(-)^n(n + 3/2 - i - j) + 1/2 (-)^{i+j} \Big] \ S_{(i|n-i)} \hat{S}_{(j|n+1-j)}
\end{equation}
The sum over colors gives the corresponding generator of the $Y(\hat{\mathfrak{gl}}_1)$:
\begin{equation}
    \hat{e}_{1}^{(1)} + \hat{e}_{1}^{(2)} = \sum_{n=1}^{\infty}  \sum_{i=1}^{n+1} \sum_{j=1}^{n} \ (-)^{i+j} \ S_{(i|n+1-i)} \hat{S}_{(j|n-j)} = \sum_{n=1}^{\infty} a \, p_{a+1} \frac{\partial}{\partial p_a}
\end{equation}
\begin{equation}
    \hat{f}_{1}^{(1)} + \hat{f}_{1}^{(2)} = \sum_{n=1}^{\infty}  \sum_{i=1}^{n+1} \sum_{j=1}^{n} \ (-)^{i+j} \ S_{(i|n-i)} \hat{S}_{(j|n+1-j)} = \sum_{n=1}^{\infty} (a+1) \, p_{a} \frac{\partial}{\partial p_{a+1}}
\end{equation}
Moreover the commutator of these sums is equal to the grading operator $\hat{\psi}_2$ from $Y(\hat{\mathfrak{gl}}_1)$ (up to a sign):
\begin{equation}\label{ef_1_mode}
    \Big[ \hat{e}_{1}^{(1)} + \hat{e}_{1}^{(2)}, \hat{f}_{1}^{(1)} + \hat{f}_{1}^{(2)} \Big] = -2 \sum_{n=1}^{\infty}  \sum_{i=1}^{n} \sum_{j=1}^{n} \ (-)^{i+j} \ S_{(i|n-i)} \hat{S}_{(j|n-j)} =  -2 \sum_{n=1}^{\infty} n \, p_{n} \frac{\partial}{\partial p_n}
\end{equation}

\subsection{$\beta$-deformation}
In this section we use the standard rule to proceed to the $\beta$-deformation:
\begin{equation}
    \epsilon_1 = 1, \hspace{10mm} \epsilon_2 = - \beta
\end{equation}
It is convenient to introduce the following notation:
\begin{align}
    \mathcal{P}_{0,n} &= \sum_{i=1}^{n} S_{(i|n-i)}, &\hspace{5mm} \widehat{\mathcal{P}}_{0,n} &= \sum_{i=1}^{n} \hat{S}_{(i|n-i)} \\
    \mathcal{P}_{1,n} &= \sum_{i=1}^{n} (n+1-2i) \, S_{(i|n-i)}, &\hspace{5mm} \widehat{\mathcal{P}}_{1,n} &= \sum_{i=1}^{n} (n+1-2i) \, \hat{S}_{(i|n-i)}
\end{align}
Applying this new notation one can rewrite the zero mode raising and lowering operators, that undergo no $\beta$-deformation:
\begin{equation}
    \hat{e}_{0}^{(1)} = p_1 + \sum_{n=1}^{\infty} (-)^n \, \mathcal{P}_{0,n+1} \widehat{\mathcal{P}}_{0,n}, \hspace{10mm} \hat{e}_{0}^{(2)} = - \sum_{n=1}^{\infty} (-)^n \, \mathcal{P}_{0,n+1} \widehat{\mathcal{P}}_{0,n}
\end{equation}

\begin{equation}
    \hat{f}_{0}^{(1)} = \frac{\partial}{\partial p_1} + \sum_{n=1}^{\infty} (-)^n \, \mathcal{P}_{0,n} \widehat{\mathcal{P}}_{0,n+1}, \hspace{10mm} \hat{f}_{0}^{(2)} = - \sum_{n=1}^{\infty} (-)^n \, \mathcal{P}_{0,n} \widehat{\mathcal{P}}_{0,n+1}
\end{equation}
Therefore the zero modes $\psi_0^{(a)}$ do not experience the $\beta$-deformation too:
\begin{equation}
    \hat{\psi}_{0}^{(1)} = -1 - 2 \sum_{n=1}^{\infty} (-)^n \, \mathcal{P}_{0,n} \widehat{\mathcal{P}}_{0,n}, \hspace{10mm} \hat{\psi}_{0}^{(2)} = 2 \sum_{n=1}^{\infty} (-)^n \, \mathcal{P}_{0,n} \widehat{\mathcal{P}}_{0,n}
\end{equation}
The fact that the zero mode operators do not depend on $\beta$ goes in parallel with the situation in $Y(\hat{\mathfrak{gl}}_1)$, where $\hat{e}_0, \hat{f}_0, \hat{\psi}_2$ operators remain undeformed \cite{MT}.
The first mode of $\hat{\psi}^{(1)}_{1}$ undergoes the following $\beta$-deformation:
\begin{tcolorbox}
\begin{equation}
\label{cut-n-join beta 1}
    \hat{\psi}^{(1)}_{1} = \frac{1-\beta}{2} -\sum_{n=1}^{\infty} (-)^n\Big[ \mathcal{P}_{0,n} \widehat{\mathcal{P}}_{1,n} + \beta \, \mathcal{P}_{1,n} \widehat{\mathcal{P}}_{0,n}  + (1-\beta) \, (n-1) \, \mathcal{P}_{0,n} \widehat{\mathcal{P}}_{0,n}   \Big]  - \frac{1-\beta}{2} \Big( \hat{\psi}^{(1)}_{0} \Big)^2
\end{equation}
\begin{equation}
\label{cut-n-join beta 2}
    \hat{\psi}^{(2)}_{1} = \frac{1-\beta}{2} +\sum_{n=1}^{\infty} (-)^n\Big[ \mathcal{P}_{0,n} \widehat{\mathcal{P}}_{1,n} + \beta \, \mathcal{P}_{1,n} \widehat{\mathcal{P}}_{0,n}  + (1-\beta) \, (n+1) \, \mathcal{P}_{0,n} \widehat{\mathcal{P}}_{0,n}   \Big]  - \frac{1-\beta}{2} \Big( \hat{\psi}^{(1)}_{0} \Big)^2
\end{equation}
\end{tcolorbox}
Note that there is no normal ordering in the above formulas including square of operator $\psi_{0}^{(1)}$. Rank two Uglov polynomials are the eigenvalues of these operators:
\begin{equation}
    \hat{\psi}_{1}^{(a)} \, U^{(2)}_{\lambda} = \psi_{\lambda, 1}^{(a)} \, U^{(2)}_{\lambda}
\end{equation}
with the following eigenvalues:
\begin{equation}
\label{cut and join 1}
    \psi_{\lambda, 1}^{(1)} = 2  \left( \kappa_{\lambda}^{(1)} - \kappa_{\lambda}^{(2)} \right) -2 (1-\beta)  \left(n^{(1)}_{\lambda} - n^{(2)}_{\lambda} \right)  \left(n^{(1)}_{\lambda} - n^{(2)}_{\lambda} - 1 \right)
\end{equation}
\begin{equation}
\label{cut and join 2}
    \psi_{\lambda, 1}^{(2)} = -2  \left( \kappa_{\lambda}^{(1)} - \kappa_{\lambda}^{(2)} \right) -2 (1-\beta)  \left(n^{(1)}_{\lambda} - n^{(2)}_{\lambda} \right)  \left(n^{(1)}_{\lambda} - n^{(2)}_{\lambda}  \right)
\end{equation}

 Using the above relations for generators from the small-set \eqref{small set} one can compute any generator of the Yangian $Y(\hat{\mathfrak{gl}}_2)$ applying rules \eqref{raising mode a a} and \eqref{raising mode a b}. However, in the $\beta$-deformed case the operators include Schur functions with more than one hook due to the square of operator $\psi_{0}^{(1)}$ in \eqref{cut-n-join beta 1}, \eqref{cut-n-join beta 2}. 
 
\bigskip
 
The first terms in the $\beta$-deformation of the \eqref{e_1_1_beta=1} and \eqref{e_1_2_beta=1} for $\hat{e}^{(a)}_{1}$ have the following form:
\begin{align}
    \begin{aligned}
        \hat{e}_{1}^{(1)} &= 2 \Big( S_{\begin{ytableau}  \ & \ & \ \end{ytableau}} \hat{S}_{\begin{ytableau}  \ & \  \end{ytableau}} -(\beta - 1) S_{\begin{ytableau}  \ \\ \ \\ \ \end{ytableau}} \hat{S}_{\begin{ytableau}  \ & \  \end{ytableau}} - \beta S_{\begin{ytableau}  \ \\ \ \\ \ \end{ytableau}} \hat{S}_{\begin{ytableau}  \ \\ \  \end{ytableau}} - \\
        & - S_{\begin{ytableau}  \ & \ & \ & \ \end{ytableau}} \hat{S}_{\begin{ytableau}  \ & \ & \ \end{ytableau}} - S_{\begin{ytableau}  \ & \ & \ & \ \end{ytableau}} \hat{S}_{\begin{ytableau}  \ & \ \\ \ \end{ytableau}} - S_{\begin{ytableau}  \ & \ & \ \\ \ \end{ytableau}} \hat{S}_{\begin{ytableau}  \ & \ & \ \end{ytableau}} + (\beta - 1) S_{\begin{ytableau}  \ & \ \\ \ \\ \ \end{ytableau}} \hat{S}_{\begin{ytableau}  \ & \ & \ \end{ytableau}} + (\beta - 1) S_{\begin{ytableau}  \ \\ \ \\ \ \\ \ \end{ytableau}} \hat{S}_{\begin{ytableau}  \ & \ & \ \end{ytableau}} + \\
        & + \beta S_{\begin{ytableau}  \ & \ \\ \ \\ \ \end{ytableau}} \hat{S}_{\begin{ytableau}  \ \\ \ \\ \ \end{ytableau}} + \beta S_{\begin{ytableau}  \ \\ \ \\ \ \\ \ \end{ytableau}} \hat{S}_{\begin{ytableau}  \ \\ \ \\ \ \end{ytableau}} + \left( 2\beta - 1 \right) S_{\begin{ytableau}  \ \\ \ \\ \ \\ \ \end{ytableau}} \hat{S}_{\begin{ytableau}  \ & \ \\ \ \end{ytableau}} - (\beta - 1) \underline{ S_{\begin{ytableau}  \ & \ \\ \ & \ \end{ytableau}} \hat{S}_{\begin{ytableau}  \ & \ \\ \ \end{ytableau}}} \Big) + \ldots = \\
        & = \left( \frac{1}{3} (1-\beta ) p_1^3+\beta  p_2 p_1+\frac{2}{3} (1-\beta ) p_3 \right) \frac{\partial^2}{\partial p_1^2} + \frac{2}{3}\left(  p_1^3+2 p_3 \right) \frac{\partial}{\partial p_2} - \frac{2}{3}\left( p_1^4+2 p_3 p_1 \right) \frac{\partial^2}{\partial p_1 \partial p_2} + \\ 
        &- \left( \frac{2}{3} (1-\beta ) p_1^4+\beta  p_2 p_1^2-\frac{2}{3} (1-\beta ) p_3 p_1-\beta  p_4 \right)\frac{\partial}{\partial p_3} +\\
        &+ \left(\frac{1}{9} (\beta -1) p_1^4-\frac{2}{3} \beta  p_2 p_1^2-\frac{8}{9} (1-\beta ) p_3 p_1-\frac{\beta  p_4}{3} \right) \frac{\partial^3}{\partial p_1^3} + \ldots
    \end{aligned}
\end{align}

\begin{align}
    \begin{aligned}
        \hat{e}_1^{(2)} &= S_{\begin{ytableau}  \ & \  \end{ytableau}} \hat{S}_{\begin{ytableau}             \         \end{ytableau}} + (1 - 2 \beta) S_{\begin{ytableau}  \ \\ \  \end{ytableau}} \hat{S}_{\begin{ytableau}             \         \end{ytableau}} -\\
        &-S_{\begin{ytableau}  \ & \ & \ \end{ytableau}} \hat{S}_{\begin{ytableau}  \ & \  \end{ytableau}} -S_{\begin{ytableau}  \ & \ & \ \end{ytableau}} \hat{S}_{\begin{ytableau}  \ \\ \  \end{ytableau}} + (2 \beta - 3) S_{\begin{ytableau}  \ & \ \\ \ \end{ytableau}} \hat{S}_{\begin{ytableau}  \ & \  \end{ytableau}} + (2\beta - 1) S_{\begin{ytableau}  \ & \ \\ \ \end{ytableau}} \hat{S}_{\begin{ytableau}  \ \\ \  \end{ytableau}} + (2\beta -1) S_{\begin{ytableau}  \ \\ \ \\ \ \end{ytableau}} \hat{S}_{\begin{ytableau}  \ & \  \end{ytableau}} + (2\beta-1) S_{\begin{ytableau}  \ \\ \ \\ \ \end{ytableau}} \hat{S}_{\begin{ytableau}  \ \\ \  \end{ytableau}} + \\
        &+ 3 S_{\begin{ytableau}  \ & \ & \ & \ \end{ytableau}} \hat{S}_{\begin{ytableau}  \ & \ & \ \end{ytableau}} + S_{\begin{ytableau}  \ & \ & \ & \ \end{ytableau}} \hat{S}_{\begin{ytableau}  \ & \ \\ \ \end{ytableau}} + S_{\begin{ytableau}  \ & \ & \ & \ \end{ytableau}} \hat{S}_{\begin{ytableau}  \ \\ \ \\ \ \end{ytableau}} + (3-2\beta) S_{\begin{ytableau}  \ & \ & \ \\ \ \end{ytableau}} \hat{S}_{\begin{ytableau}  \ & \ & \ \end{ytableau}} +  (3-2\beta)S_{\begin{ytableau}  \ & \ & \ \\ \ \end{ytableau}} \hat{S}_{\begin{ytableau}  \ & \ \\ \ \end{ytableau}} + (1-2\beta) S_{\begin{ytableau}  \ & \ & \ \\ \ \end{ytableau}} \hat{S}_{\begin{ytableau}  \ \\ \ \\ \ \end{ytableau}} + \\
        &+(3-2\beta) S_{\begin{ytableau}  \ & \ \\ \ \\ \ \end{ytableau}} \hat{S}_{\begin{ytableau}  \ & \ & \ \end{ytableau}} +  (3-4\beta)S_{\begin{ytableau}  \ & \ \\ \ \\ \ \end{ytableau}} \hat{S}_{\begin{ytableau}  \ & \ \\ \ \end{ytableau}} + (1-2\beta) S_{\begin{ytableau}  \ & \ \\ \ \\ \ \end{ytableau}} \hat{S}_{\begin{ytableau}  \ \\ \ \\ \ \end{ytableau}} +(3-4\beta) S_{\begin{ytableau}  \ \\ \ \\ \ \\ \ \end{ytableau}} \hat{S}_{\begin{ytableau}  \ & \ & \ \end{ytableau}} +  (1-2\beta)S_{\begin{ytableau}  \ \\ \ \\ \ \\ \ \end{ytableau}} \hat{S}_{\begin{ytableau}  \ & \ \\ \ \end{ytableau}} + \\
        &+(1-4\beta) S_{\begin{ytableau}  \ \\ \ \\ \ \\ \ \end{ytableau}} \hat{S}_{\begin{ytableau}  \ \\ \ \\ \ \end{ytableau}}+2(1-\beta) \underline{S_{\begin{ytableau}  \ & \ \\ \ & \ \end{ytableau}} \hat{S}_{\begin{ytableau}  \ & \ \\ \ \end{ytableau}}} + \ldots  = \\
        &= \left( (1-\beta) p_1^2 + \beta p_2 \right) \frac{\partial}{\partial p_1} + \left( (\beta -1) p_1^3-\beta  p_1 p_2 \right) \frac{\partial^2}{\partial p_1^2} -\frac{2}{3} \left(p_1^3-p_3\right) \frac{\partial}{\partial p_2} + \frac{2}{3} \left(p_1^4+2 p_3 p_1\right) \frac{\partial^2}{\partial p_1 \partial p_2} + \\ 
        &+\frac{1}{9} \left(5 (1-\beta ) p_1^4+6 \beta  p_2 p_1^2-4 (\beta -1) p_3 p_1+3 \beta  p_4\right) \frac{\partial^3}{\partial p_1^3} +\\
        &+ \left( -\frac{1}{3} (\beta -1) p_1^4+\beta  p_2 p_1^2-\frac{8}{3} (\beta -1) p_3 p_1+2 \beta  p_4 \right) \frac{\partial}{\partial p_3} + \ldots 
    \end{aligned}
\end{align}
The underlined term in the above formula correspond to the first 2-hook contribution.

The $\beta$-deformation of the $Y(\widehat{\fg\fl}_1)$ generator \eqref{ef_1_mode} is
\begin{align}
    \begin{aligned}
        \hat{e}_{1}^{(1)} + \hat{e}_{1}^{(2)} &= \beta p_2 \frac{\partial}{ \partial p_1} + 2 p_3 \frac{\partial}{ \partial p_2}+ 3\beta p_4 \frac{\partial}{\partial p_3} + \ldots \\
        &+(1 - \beta) \Bigg\{ p_1^2 \frac{\partial}{ \partial p_1} +\frac{2}{3} \left(p_3-p_1^3\right) +\frac{\partial^2}{\partial p_1^2} + \frac{4}{9} \left(p_1^4-p_1 p_3\right) \frac{\partial^3}{\partial p_1^3} + \frac{1}{3} \left(10 p_1 p_3-p_1^4\right) \frac{\partial}{\partial p_3} + \ldots \Bigg\}
    \end{aligned}
\end{align}
Note that the last correcting piece is non-trivial, however it does not depend on even times and derivatives. This fact is tightly related with the absence of the order-two operator in the commuting family of operators from Section \ref{sec:time_op}.
 
The above formulas for Yangian operators force the following normalization for rank 2 Uglov polynomials reads:
\begin{align}
    \begin{aligned}
    \label{Uglovs}
        U_{ \varnothing } &= 1, &\hspace{10mm} U_{\begin{ytableau} \ \end{ytableau}} &= p_1 \\
        U_{\begin{ytableau}  \ & \  \end{ytableau}} &= -\left(\beta  p_2+p_1^2 \right), &\hspace{10mm} U_{\begin{ytableau}  \ \\ \  \end{ytableau}} &= \beta  \left(p_1^2-p_2\right) \\
        U_{\begin{ytableau}  \ & \ & \ \end{ytableau}} &= -\frac{1}{3} \left(3 \beta  p_2 p_1+p_1^3+2 p_3\right), &\hspace{10mm} U_{\begin{ytableau}  \ \\ \ \\ \ \end{ytableau}} &= \frac{1}{3} \beta  \left(p_1^3-3 p_2 p_1+2 p_3\right) \\
        U_{\begin{ytableau}  \ & \ \\ \ \end{ytableau}} &= \frac{1}{3} \left(p_1^3-p_3\right) \\
        \ldots
    \end{aligned}
\end{align}
\begin{comment}
\begin{align}
    \begin{aligned}
        U_{\begin{ytableau}  \ & \ & \ & \ \end{ytableau}} &= -\frac{1}{3} \left(6 \beta  p_2 p_1^2+3 \beta  \left(\beta  p_2^2+2 p_4\right)+p_1^4+8 p_3 p_1\right) \\
        U_{\begin{ytableau}  \ & \ & \ \\ \ \end{ytableau}} &= \frac{1}{3} \left((\beta +2) p_1^4+3 \beta  (\beta +1) p_2 p_1^2+2 (\beta -1) p_3 p_1-3 \beta  \left(\beta  p_2^2+2 p_4\right)\right) \\
        U_{\begin{ytableau}  \ & \ \\ \ & \ \end{ytableau}} &= -\frac{1}{3} \beta  \left(3 \beta  p_2^2-3 (\beta -1) p_4+p_1^4-4 p_3 p_1\right) \\
        U_{\begin{ytableau}  \ & \ \\ \ \\ \ \end{ytableau}} &=\frac{1}{3} \beta  \left((2 \beta +1) p_1^4-3 (\beta +1) p_2 p_1^2-2 (\beta -1) p_3 p_1-3 \beta  \left(p_2^2-2 p_4\right)\right) \\
        U_{\begin{ytableau}  \ \\ \ \\ \ \\ \ \end{ytableau}} &=-\frac{1}{3} \beta ^2 \left(p_1^4-6 p_2 p_1^2+8 p_3 p_1+3 p_2^2-6 p_4\right)
    \end{aligned}
\end{align}
\end{comment}
we dropped for simplicity the rank $U^{(2)}_{\lambda} = U_{\lambda}$. The action of raising operators on Uglov polynomials:
\begin{align}
    \begin{aligned}
        &\hat{e}_{k}^{(1)} \, U_{\varnothing} = \delta_{k,0} \, U_{\begin{ytableau}
            \
        \end{ytableau}}, &\hspace{10mm} &\hat{e}_{k}^{(2)} \, U_{\varnothing} = 0 \\
        &\hat{e}_{k}^{(1)} \, U_{\begin{ytableau} \ \end{ytableau}} = 0, &\hspace{10mm} &\hat{e}_{k}^{(2)} \, U_{\begin{ytableau} \ \end{ytableau}} = \frac{(-1)}{1+\beta} \, U_{\begin{ytableau}  \ & \  \end{ytableau}} + \frac{(-\beta)^k}{1+\beta} \, U_{\begin{ytableau}  \ \\ \  \end{ytableau}} \\
        &\hat{e}_{k}^{(1)} \, U_{\begin{ytableau}  \ & \  \end{ytableau}} = 2^k \, U_{\begin{ytableau}  \ & \ & \ \end{ytableau}}, &\hspace{10mm} &\hat{e}_{k}^{(2)} \, U_{\begin{ytableau}  \ & \  \end{ytableau}} = -2(-\beta)^k \, U_{\begin{ytableau}  \ & \ \\ \ \end{ytableau}} \\
        &\hat{e}_{k}^{(1)} \, U_{\begin{ytableau}  \ \\ \  \end{ytableau}} = (-2\beta)^k \, U_{\begin{ytableau}  \ \\ \ \\ \ \end{ytableau}}, &\hspace{10mm} &\hat{e}_{k}^{(2)} \, U_{\begin{ytableau}  \ \\ \  \end{ytableau}} = 2\beta \, U_{\begin{ytableau}  \ & \ \\ \ \end{ytableau}} \\
        &\hat{e}_{k}^{(1)} U_{\begin{ytableau}  \ & \ & \ \end{ytableau}} = 0, &\hspace{10mm} &\hat{e}_{k}^{(2)} U_{\begin{ytableau}  \ & \ & \ \end{ytableau}} = \frac{3^k}{\beta + 3} \, U_{\begin{ytableau}  \ & \ & \ & \ \end{ytableau}} - \frac{(-\beta)^k}{\beta + 3} \, U_{\begin{ytableau}  \ & \ & \ \\ \ \end{ytableau}} \\
        &\hat{e}_{k}^{(1)} U_{\begin{ytableau}  \ \\ \ \\ \ \end{ytableau}} = 0, &\hspace{10mm} &\hat{e}_{k}^{(2)} U_{\begin{ytableau}  \ \\ \ \\ \ \end{ytableau}} = -\frac{(-3\beta)^k}{1+3\beta} \, U_{\begin{ytableau}  \ \\ \ \\ \ \\ \ \end{ytableau}} +\frac{1}{1+3\beta} \, U_{\begin{ytableau}  \ & \ \\ \ \\ \ \end{ytableau}} \\
        \ldots
    \end{aligned}
\end{align}
\begin{comment}
\begin{equation}
    \hat{e}_{k}^{(1)} \, U_{\begin{ytableau}  \ & \ \\ \ \end{ytableau}} = \frac{2^k}{2(1+\beta)^2} \, U_{\begin{ytableau}  \ & \ & \ \\ \ \end{ytableau}} - \frac{(1-\beta)^k}{(1+\beta)^2} \, U_{\begin{ytableau}  \ & \ \\ \ & \ \end{ytableau}} + \frac{(-2\beta)^k}{2(1+\beta)^2} \, U_{\begin{ytableau}  \ & \ \\ \ \\ \ \end{ytableau}} , \hspace{10mm} \hat{e}_{k}^{(2)} \, U_{\begin{ytableau}  \ & \ \\ \ \end{ytableau}} = 0
\end{equation}
\end{comment}
The apparent asymmetry between $\hat e_k^{(1)}$ and $\hat e_k^{(2)}$ is obviously related to the chess-coloring of the diagrams. For example, the ``first'' operator $\hat e_k^{(1)}$ annihilates the single box of $U_{\begin{ytableau}  \ \end{ytableau}}$, thus it can not contain
derivative $\frac{\p}{\p p_1}$ -- only the ``second'' operator $\hat e_k^{(2)}$ does -- and so on.
For example, $U_{\begin{ytableau}  \ & \ \\ \ \end{ytableau}}$ is annihilated by $\hat e_k^{(2)}$ -- but not $U_{\begin{ytableau}  \ & \ \end{ytableau}}$ or $U_{\begin{ytableau}  \ \\ \  \end{ytableau}}$, where both operators
act by gluing the ``first'' and the ``second'' box respectively.

Note that the normalization of 2-Uglov polynomials \eqref{Uglovs} does not match the normalization of Schur polynomials in the limit $\beta = 1$. This is the price for simple form of operators $\hat{e}^{(a)}_0$, $\hat{f}^{(a)}_0$ that do not depend on $\beta$ even in the case of $\beta$-deformed representation.

\section{Conclusion}\label{sec:conc}

The main goal of this paper was to describe the simple representation of affine Yangian algebras in terms
of time variables.
In this case the states are labeled by the ordinary Young diagrams and  represented by orthogonal polynomials
from the Schur-Macdonald family.
We have explained that for $Y(\widehat{\fg\fl}_r)$ the relevant orthogonal polynomials were actually $r$-Uglov polynomials,
of which Jacks at $r=1$ the well-known example.
The diagrams acquire additional grading or coloring in $r$ different colors,
for $r=2$ were the diagram coloring is similar to that of a chess board.
We have provided explicit expressions (``bosonisation rules'') for the small-set (\ref{small set})
of Yangian generators for $r=2$ which look suggestive enough for generalization to arbitrary $r$
and further to supersymmetric   $Y(\widehat{\fg\fl}_{m|n})$.
In the latter case the set of time variables should be extended to involve superpartners,
and the diagrams should acquire triangle fragment pieces -- as explained in \cite{Galakhov:2023mak}.
 
We have begun our story from the abstract commutation relations, implied by the quiver construction of \cite{Li:2020rij,Galakhov:2023gjs},
and demonstrated that Uglov polynomials provided their time (free-field) representation {\it a la}
\cite{MOROZOV1989239,GMMOS, Morozov:2021hwr, Morozov:2022ocp, FeFr}.
We have completed  Sect.\ref{sec: time var reps} by the inverse construction -- given a set of time-dependent orthogonal polynomials,
labeled by the Young diagrams,
we can define an algebra of differential operators, which act between them.
In fact, there is a whole family of algebras $Y(\widehat{\fg\fl}_r)$,
defined by diagonal coloring of the diagrams in $r$ colors:
one can separate operators acting between different pairs of adjacent colors.
For all $r$ these algebras can be considered as acting on the same family of Schur polynomials,
however the $\beta$-deformation splits them into different $r$-Uglov families.
These dual definitions of the algebra and its free-field representation needs thorough consideration
to be presented elsewhere.

\section*{Acknowledgements}

This work was supported by the Russian Science Foundation (Grant No.20-12-00195).

\appendix

\section{$r$-Uglov polynomials as orthogonal polynomials}\label{sec:Uglov_def}

In this appendix we describe the direct construction of Uglov polynomials by orthogonalization method
(reviewed in \cite{Mironov:2018nie}, without a direct reference to Macdonald polynomials.

For partitions we define a partial ordering.
We say that $\lambda\leq \mu$ if $|\lambda|=|\mu|$ and
\begin{equation}
	\lambda_1+\lambda_2+\ldots+\lambda_i\leq \mu_1+\mu_2+\ldots+\mu_i,\quad \forall\,i\,.
\end{equation}

Monomial symmetric functions $m_{\lambda}$ are canonically defined in the following way:
\begin{equation}
	m_{\lambda}(x_1,\ldots,x_n)=\frac{1}{\left|\left\{\sigma\,|\,\sigma\in S_n,\,\sigma\cdot\lambda=\lambda \right\}\right|}\sum\lm_{\sigma\in S_n}x_1^{\lambda_{\sigma(1)}}x_2^{\lambda_{\sigma(2)}}\ldots x_n^{\lambda_{\sigma(n)}}\,.
\end{equation}

Times $p_k$ coincide with selected symmetric combinations:
\begin{equation}
	p_k:=m_{[k]},\quad p_{\lambda}:=\prod\lm_ip_{\lambda_i}\,.
\end{equation}

There is a canonical $r$-norm on polynomials:
\begin{equation}
	\left\langle p_{\lambda},p_{\mu}\right\rangle_r=\delta_{\mu,\nu}z_{\lambda}\beta^{|\{i|\lambda_i=0\,{\rm mod}\,r\}|}\,,
\end{equation}
where $z=\prod\lm_{k=1}^{\infty}k^{u_k}u_k!$ for $\lambda=(1^{u_1},2^{u_2},\ldots)$ is the standard norm for Schur polynomials.
Apparently, with respect to this norm $p_k$ is conjugated to a differential operator
\begin{equation}
	p_k^*=\beta^{\delta_{0,(k\; {\rm mod}\; r)}}k\frac{\p}{\p p_k}\,.
\end{equation}

\emph{Unnormalized} $r$-Uglov polynomials ${\rm Pol}_{\lambda}^{(r)}$ are defined uniquely by the following rules:
\begin{enumerate}
	\item ${\rm Pol}_{\lambda}^{(r)}=m_{\lambda}+\sum\lm_{\mu < \lambda}u_{\lambda\mu}m_{\mu}$ for some $u_{\lambda\mu}$.
	\item $\left\langle {\rm Pol}_{\lambda}^{(r)},{\rm Pol}_{\mu}^{(r)}\right\rangle_r=0$ for $\lambda\neq \mu$
\end{enumerate}

Define \emph{normalized} $r$-Uglov polynomials $U_{\lambda}^{(r)}$ in the following way:
\begin{equation}
	U_{\lambda}^{(r)}:={\rm Pol}_{\lambda}^{(r)}\times \prod\lm_{\Box\in\lambda_r}\left(\left({\rm arm}_{\lambda}(\Box)+1\right)+\beta\,{\rm leg}_{\lambda}(\Box)\right)\,,
\end{equation}
where
\begin{equation}
	\lambda_r:=\left\{\Box\in\lambda\,\big|\,{\rm leg}_{\lambda}(\Box)+{\rm arm}_{\lambda}(\Box)+1=0\;{\rm mod}\;r\right\}\,.
\end{equation}

We observe that the norm of $r$-Uglov polynomials:
\begin{equation}
	\left\langle U^{(r)}_{\lambda},U^{(r)}_{\mu}\right\rangle_r=\delta_{\lambda,\mu}\CN^{(r)}_{\lambda}
\end{equation}
reads
\begin{equation}
	\CN_{\lambda}^{(r)}=\prod\lm_{\Box\in\lambda_r}\left(\left({\rm arm}_{\lambda}(\Box)+1\right)+\beta{\rm leg}_{\lambda}(\Box)\right)\left({\rm arm}_{\lambda}(\Box)+\beta\left({\rm leg}_{\lambda}(\Box)+1\right)\right)\,.
\end{equation}

\section{Hook formulas from LMNS integrals} \label{sec:LMNS}

In this appendix we outline the possibility to derive the hook formulas for the Euler classes, representing a canonical normalization of BPS wavefunctions,
from the Losev-Moore-Nekrasov-Shatashvili (LMNS) formalism \cite{Losev:1995cr,Moore:1997dj,Moore:1998et} (see also \cite{Kimura:2015rgi,Pestun:2016zxk,Bao:2024noj} for recent reviews).
This is not truly important for the practical applications of bosonisation,
i.e. representations through the time variables.
However we expect that this analysis might be  easily extended to the case of affine super Yangians $Y(\widehat{\fg\fl}_{m|n})$ where we expect simple natural expressions for the (semi-)Fock representations of D-brane states on smooth 4-cycles in CY${}_3$'s in the LMNS formalism, and generic combinatorial expressions counting hooks are unknown.
An explicit example of the hook formulas for the case of $Y(\widehat{\fg\fl}_{1|1})$ was constructed in \cite{Galakhov:2023mak}.

Consider a quiver variety spanned by the chiral fields $\mathscr{F}$ defined (as an algebraic variety) as solutions to a set of equations $\mathscr{E}$.
Also there is an action of the gauge group $\mathscr{G}$.
If this variety is smooth\footnote{
	We should remind that this strategy is applicable only to smooth quiver varieties.
	Fock representations of $Y(\widehat{\fg\fl}_r)$ are described by Nakajima quiver varieties that are known to be smooth \cite{2009arXiv0905.0686G}.
	A notorious counter-example for this set-up is a generic MacMahon module \cite{Morozov:2022ndt} (plain partitions) of $Y(\widehat{\fg\fl}_1)$ corresponding to Hilbert scheme ${\rm Hilb}^*(\IC^3)$ \cite{Okounkov:2015spn}.
} the corresponding Euler class of the tangent bundle to fixed point $\lambda$ (Young diagram) can be found with the help of the Losev-Moore-Nekrasov-Shatashvili (LMNS) residue formula:
\begin{equation}
	{\rm Eul}_{\lambda}^{-1}\;=\;\mathop{\rm res}\lm_{\lambda}\,\frac{w(\mathscr{G})\times w(\mathscr{E})}{w(\mathscr{F})}\,,
\end{equation}
where we calculate the weights as eigenvalues of operator ${\rm Ad}_\Phi-\epsilon$, where matrix $\Phi$ is diagonal yet has unfixed eigenvalues $\Phi={\rm diag}(\phi_1,\ldots,\phi_n)$.
The residue (also known as the Jeffrey-Kirwan residue \cite{JEFFREY1995291}) is taken in the point where $\phi_i$'s acquire values geometrically corresponding to the boxes of diagram $\lambda$ in the $\epsilon$-plane.

For example, in the case of ADHM $Y(\widehat{\fg\fl}_1)$:
\begin{equation}
	\begin{aligned}
		&w_{ij}(B_a)=\phi_i-\phi_j-\epsilon_a\,,\\
		&w_{ij}(I)=\phi_i\,,\\
		&w_{ij}(J)=-\phi_j-\epsilon_1-\epsilon_2\,,\\
		&w_{ij}({\rm gauge})=\phi_i-\phi_j\,,\\
		&w_{ij}([B_1,B_2]-IJ)=-\left(\phi_i-\phi_j+\epsilon_1+\epsilon_2\right)\,.
	\end{aligned}
\end{equation}

Therefore we get the LMNS formula:
\begin{equation}
	{\rm Eul}_{\lambda}^{-1}\;=\;\oint\prod\lm_id\phi_i\times\prod\lm_{i<j}u(\phi_i-\phi_j)\times\prod\lm_{i}\frac{1}{\phi_i(\phi_i+\epsilon_1+\epsilon_2)}\,,
\end{equation}
where
\begin{equation}
	u(z):=\frac{z^2\left(z^2-(\epsilon_1+\epsilon_2)^2\right)}{\left(z^2-\epsilon_1^2\right)\left(z^2-\epsilon_2^2\right)}\,.
\end{equation}

The LMNS formula proposes the following recurrent relations for the Euler classes:
\begin{equation}\label{recurrent}
	{\rm Eul}_{\lambda+\Box}^{-1}={\rm Eul}_{\lambda}^{-1}\times\mathop{\rm res}\lm_{z=x_{\Box}}\prod\lm_{\Box'\in\lambda}u(z-x_{\Box'})\times\frac{1}{z(z+\epsilon_1+\epsilon_2)}\,.
\end{equation}

Note that function $u$ has nice "\emph{cocyclish}" properties:
\begin{equation}
	\prod\lm_{k=0}^{n}u(z-k\epsilon_1)=\frac{z(z-n\epsilon_1)(z-(n+1)\epsilon_1-\epsilon_2)(z+\epsilon_1+\epsilon_2)}{(z-(n+1)\epsilon_1)(z+\epsilon_1)(z-\epsilon_2)(z-n\epsilon_1+\epsilon_2)}\,.
\end{equation}

Let us introduce the following notation for a potential of area $\mathscr{D}$ inside the diagram:
\begin{equation}
	U_z(\mathscr{D}):=\prod\lm_{\Box\in\mathscr{D}}u(z-x_{\Box})\,.
\end{equation}

For a horizontal strip
\begin{equation}
	L(x_1,x_2)=\begin{array}{c}
		\begin{tikzpicture}[scale=0.2]
			\foreach \i/\j in {0/-1, 0/0, 1/-1, 1/0, 2/-1, 2/0, 3/-1, 3/0, 4/-1, 4/0, 5/-1, 5/0, 6/-1, 6/0, 7/-1, 7/0, 8/-1, 8/0}
			{
				\draw (\i,\j) -- (\i+1,\j);
			}
			\foreach \i/\j in {0/0, 1/0, 2/0, 3/0, 4/0, 5/0, 6/0, 7/0, 8/0, 9/0}
			{
				\draw (\i,\j) -- (\i,\j-1);
			}
			\node[left] at (0,-0.5) {$\scriptstyle x_1$};
			\node[left] at (12,-0.5) {$\scriptstyle x_2$};
		\end{tikzpicture}
	\end{array}
\end{equation}
we have:
\begin{equation}\label{horiz}
	U_z[L(x_1,x_2)]=\frac{(z-x_1)(z-x_1+\epsilon_1+\epsilon_2)(z-x_2)(z-x_2-\epsilon_1-\epsilon_2)}{(z-x_1+\epsilon_1)(z-x_1-\epsilon_2)(z-x_2-\epsilon_1)(z-x_2+\epsilon_2)}\,.
\end{equation}
Actually, $u$-function is a 2-cocycle, and for a parallelogram we have:
\begin{equation}\label{parallelogram}
	U_z\left[\begin{array}{c}
		\begin{tikzpicture}[scale=0.2,yscale=-1]
			\foreach \i/\j in {0/-3, 0/-2, 0/-1, 0/0, 1/-3, 1/-2, 1/-1, 1/0, 2/-3, 2/-2, 2/-1, 2/0, 3/-3, 3/-2, 3/-1, 3/0}
			{
				\draw (\i,\j) -- (\i+1,\j);
			}
			\foreach \i/\j in {0/-2, 0/-1, 0/0, 1/-2, 1/-1, 1/0, 2/-2, 2/-1, 2/0, 3/-2, 3/-1, 3/0, 4/-2, 4/-1, 4/0}
			{
				\draw (\i,\j) -- (\i,\j-1);
			}
			\node[above left] at (0,-3) {$\scriptstyle x_1$};
			\node[below left] at (0,0) {$\scriptstyle x_3$};
			\node[above right] at (4,-3) {$\scriptstyle x_2$};
			\node[below right] at (4,0) {$\scriptstyle x_4$};
		\end{tikzpicture}
	\end{array}\right]=\frac{(z-x_1)(z-x_1+\epsilon_1+\epsilon_2)(z-x_4)(z-x_4-\epsilon_1-\epsilon_2)}{(z-x_2-\epsilon_1)(z-x_2+\epsilon_2)(z-x_3+\epsilon_1)(z-x_3-\epsilon_2)}\,.
\end{equation}

As we have seen in \eqref{horiz} the function $U_z$ is factorized:
\begin{equation}
	U_z[L(x_1,x_2)]=l(x_1)r(x_2)\,.
\end{equation}
Representing a parallelogram as a composition of horizontal strips we note that factors $l(x_1)$ and $r(x_2)$ cancel in the parallelogram product separately.
This observation allows one to derive immediately a product formula for a trapezoid with one uneven side:
\begin{equation}\label{trapezoid1}
	U_z\left[\begin{array}{c}
		\begin{tikzpicture}[scale=0.3]
			\foreach \i/\j in {0/-4, 0/-3, 0/-2, 0/-1, 0/0, 1/-4, 1/-3, 1/-2, 1/-1, 1/0, 2/-4, 2/-3, 2/-2, 2/-1, 2/0, 3/-3, 3/-2, 3/-1, 3/0, 4/-3, 4/-2, 4/-1, 4/0, 5/-2, 5/-1, 5/0, 6/-1, 6/0, 7/-1, 7/0}
			{
				\draw (\i,\j) -- (\i+1,\j);
			}
			\foreach \i/\j in {0/-3, 0/-2, 0/-1, 0/0, 1/-3, 1/-2, 1/-1, 1/0, 2/-3, 2/-2, 2/-1, 2/0, 3/-3, 3/-2, 3/-1, 3/0, 4/-2, 4/-1, 4/0, 5/-2, 5/-1, 5/0, 6/-1, 6/0, 7/0, 8/0}
			{
				\draw (\i,\j) -- (\i,\j-1);
			}
			\node[above left] at (0,0) {$\scriptstyle x_1$};
			\node[below left] at (0,-4) {$\scriptstyle x_3$};
			\node[right] at (8,-0.5) {$\scriptstyle y_1$};
			\node[right] at (6,-1.5) {$\scriptstyle y_2$};
			\node[right] at (5,-2.5) {$\scriptstyle y_3$};
			\node[right] at (3,-3.5) {$\scriptstyle y_4$};
		\end{tikzpicture}
	\end{array}\right]=\frac{(z-x_1)(z-x_1+\epsilon_1+\epsilon_2)}{(z-x_3+\epsilon_1)(z-x_3-\epsilon_2)}\times\prod\lm_{k=1}^n\frac{(z-y_k)(z-y_k-\epsilon_1-\epsilon_2)}{(z-y_k-\epsilon_1)(z-y_k+\epsilon_2)}\,.
\end{equation}
Similarly, we derive:
\begin{equation}\label{trapezoid2}
	U_z\left[\begin{array}{c}
		\begin{tikzpicture}[yscale=-1,scale=0.3,rotate=90]
			\foreach \i/\j in {0/-4, 0/-3, 0/-2, 0/-1, 0/0, 1/-4, 1/-3, 1/-2, 1/-1, 1/0, 2/-4, 2/-3, 2/-2, 2/-1, 2/0, 3/-3, 3/-2, 3/-1, 3/0, 4/-3, 4/-2, 4/-1, 4/0, 5/-2, 5/-1, 5/0, 6/-1, 6/0, 7/-1, 7/0}
			{
				\draw (\i,\j) -- (\i+1,\j);
			}
			\foreach \i/\j in {0/-3, 0/-2, 0/-1, 0/0, 1/-3, 1/-2, 1/-1, 1/0, 2/-3, 2/-2, 2/-1, 2/0, 3/-3, 3/-2, 3/-1, 3/0, 4/-2, 4/-1, 4/0, 5/-2, 5/-1, 5/0, 6/-1, 6/0, 7/0, 8/0}
			{
				\draw (\i,\j) -- (\i,\j-1);
			}
			\node[above left] at (0,0) {$\scriptstyle x_1$};
			\node[above right] at (0,-4) {$\scriptstyle x_2$};
			\node[below] at (8,-0.5) {$\scriptstyle y_1$};
			\node[below] at (6,-1.5) {$\scriptstyle y_2$};
			\node[below] at (5,-2.5) {$\scriptstyle y_3$};
			\node[below] at (3,-3.5) {$\scriptstyle y_4$};
		\end{tikzpicture}
	\end{array}\right]=\frac{(z-x_1)(z-x_1+\epsilon_1+\epsilon_2)}{(z-x_2-\epsilon_1)(z-x_2+\epsilon_2)}\times\prod\lm_{k=1}^n\frac{(z-y_k)(z-y_k-\epsilon_1-\epsilon_2)}{(z-y_k+\epsilon_1)(z-y_k-\epsilon_2)}\,.
\end{equation}

For a generic diagram we naturally divide it in three area $T_1$, $T_2$, $T_3$ (see Fig.~\ref{fig:diagrams}(a)) and integrate potential over them using \eqref{parallelogram}, \eqref{trapezoid1} and \eqref{trapezoid2} respectively.
Due to the cocycle identities these products cancel each other except pairwise contributions of the new box and the boxes at the diagram surface to the left and to the right (see Fig.~\ref{fig:diagrams}(b)).

\begin{figure}[ht!]
	\begin{center}
		\begin{tikzpicture}[scale=0.35]
			\begin{scope}
				\draw[ultra thin,fill=gray] (0,0) -- (4,0) -- (4,-3) -- (0,-3) -- cycle;
				\draw[ultra thin,fill=burgundy] (0,-3) -- (4,-3) -- (4,-4) -- (2,-4) -- (2,-5) -- (0,-5) -- cycle;
				\draw[ultra thin,fill=white!40!blue] (4,0) -- (4,-3) -- (6,-3) -- (6,-1) -- (8,-1) -- (8,0) -- cycle;
				\draw[ultra thin,fill=orange] (4,-3) -- (4,-4) -- (5,-4) -- (5,-3) -- cycle;
				\foreach \i/\j in {0/-5, 0/-4, 0/-3, 0/-2, 0/-1, 0/0, 1/-5, 1/-4, 1/-3, 1/-2, 1/-1, 1/0, 2/-4, 2/-3, 2/-2, 2/-1, 2/0, 3/-4, 3/-3, 3/-2, 3/-1, 3/0, 4/-4, 4/-3, 4/-2, 4/-1, 4/0, 5/-3, 5/-2, 5/-1, 5/0, 6/-1, 6/0, 7/-1, 7/0}
				{
					\draw[thick] (\i,\j) -- (\i+1,\j);
				}
				\foreach \i/\j in {0/-4, 0/-3, 0/-2, 0/-1, 0/0, 1/-4, 1/-3, 1/-2, 1/-1, 1/0, 2/-4, 2/-3, 2/-2, 2/-1, 2/0, 3/-3, 3/-2, 3/-1, 3/0, 4/-3, 4/-2, 4/-1, 4/0, 5/-3, 5/-2, 5/-1, 5/0, 6/-2, 6/-1, 6/0, 7/0, 8/0}
				{
					\draw[thick] (\i,\j) -- (\i,\j-1);
				}
				\node[above left, gray] at (0,0) {$\scriptstyle T_1$};
				\node[below, burgundy] at (1,-5) {$\scriptstyle T_3$};
				\node[right, white!40!blue] at (6,-2) {$\scriptstyle T_2$};
				\node[below right, orange] at (5,-4) {$\scriptstyle z$};
			\end{scope}
			\begin{scope}[shift = {(15,0)}]
				\draw[ultra thin,fill=gray] (1,-5) -- (1,-3) -- (4,-3) -- (4,0) -- (8,0) -- (8,-1) -- (5,-1) -- (5,-4) -- (2,-4) -- (2,-5) -- cycle;
				\draw[ultra thin,fill=orange] (4,-3) -- (4,-4) -- (5,-4) -- (5,-3) -- cycle;
				\draw[ultra thin,fill=burgundy] (1,-3) -- (1,-4) -- (2,-4) -- (2,-3) -- cycle;
				\draw[ultra thin,fill=white!40!blue] (4,0) -- (4,-1) -- (5,-1) -- (5,0) -- cycle;
				\foreach \i/\j in {0/-5, 0/-4, 0/-3, 0/-2, 0/-1, 0/0, 1/-5, 1/-4, 1/-3, 1/-2, 1/-1, 1/0, 2/-4, 2/-3, 2/-2, 2/-1, 2/0, 3/-4, 3/-3, 3/-2, 3/-1, 3/0, 4/-4, 4/-3, 4/-2, 4/-1, 4/0, 5/-3, 5/-2, 5/-1, 5/0, 6/-1, 6/0, 7/-1, 7/0}
				{
					\draw[thick] (\i,\j) -- (\i+1,\j);
				}
				\foreach \i/\j in {0/-4, 0/-3, 0/-2, 0/-1, 0/0, 1/-4, 1/-3, 1/-2, 1/-1, 1/0, 2/-4, 2/-3, 2/-2, 2/-1, 2/0, 3/-3, 3/-2, 3/-1, 3/0, 4/-3, 4/-2, 4/-1, 4/0, 5/-3, 5/-2, 5/-1, 5/0, 6/-2, 6/-1, 6/0, 7/0, 8/0}
				{
					\draw[thick] (\i,\j) -- (\i,\j-1);
				}
				\node[below right, orange] at (5,-4) {$\scriptstyle z$};
				\node[below] at (1.5,-5) {$\scriptstyle y_L$};
				\node[right] at (8,-0.5) {$\scriptstyle y_R$};
			\end{scope}
			\node at (3,1.5) {(a)};
			\node at (18,1.5) {(b)};
		\end{tikzpicture}
		\caption{Cutting Young diagrams into hooks.}\label{fig:diagrams}
	\end{center}
\end{figure}

Note that the mutual coordinate differences of the boundary boxes and the new box can be rewritten in terms of hooks:
\begin{equation}
	\begin{aligned}
		&z-y_L=\epsilon_1 h_{\color{burgundy}\blacksquare}-\epsilon_2 v_{\color{burgundy}\blacksquare}\,,\\
		&z-y_R=-\epsilon_1 h_{\color{white!40!blue}\blacksquare}+\epsilon_2 v_{\color{white!40!blue}\blacksquare}\,,
	\end{aligned}
\end{equation}
where $h$ and $v$ functions represent the numbers of boxes in the horizontal and vertical straight strips to the right and to the bottom of the hook junction box.

Hence for the left and right contributions we derive:
\begin{equation}
	\frac{(z-y_L)(z-y_L-\epsilon_1-\epsilon_2)}{(z-y_L+\epsilon_1)(z-y_L-\epsilon_2)}=\frac{{\rm hook}_{\lambda}({\color{burgundy}\blacksquare})}{{\rm hook}_{\lambda+\Box}({\color{burgundy}\blacksquare})},\quad \frac{(z-y_L)(z-y_L-\epsilon_1-\epsilon_2)}{(z-y_L-\epsilon_1)(z-y_L+\epsilon_2)}=\frac{{\rm hook}_{\lambda}({\color{white!40!blue}\blacksquare})}{{\rm hook}_{\lambda+\Box}({\color{white!40!blue}\blacksquare})}\,,
\end{equation}
where we introduced a hook function:
\begin{equation}
	{\rm hook}_{\lambda}(\Box):=\left(\epsilon_1(h_{\Box}+1)-\epsilon_2v_{\Box}\right)\left(\epsilon_1h_{\Box}-\epsilon_2(v_{\Box}+1)\right)\,.
\end{equation}

Then we find the following solution to the recurrent relation:
\begin{equation}
	{\rm Eul}_{\lambda}=\prod\lm_{\Box\in\lambda}{\rm hook}_{\lambda}(\Box)\,.
\end{equation}

\section{Lifting to DIM} \label{sec:DIM}

In this appendix we stress the possibility to embed the entire consideration of affine Yangian representations
into that of DIM algebras -- which in many aspects can be much simpler,
since it restores the negative modes, eliminated in the Yangian limit,
and makes the symmetry looking closer to the Lie-algebraic ones.

Macdonald polynomials   enter our framework quite nicely as a representation of the Ding-Iohara-Miki (DIM) algebra \cite{Ding:1996mq, Miki:2007mer, Mironov:2016yue, Awata:2016riz,Awata:2017lqa}, or the toroidal quantum algebra $T(\widehat{\fg\fl}_1)$.\footnote{
In terms of the rational-trigonometric-elliptic deformation hierarchy of algebras $T(\widehat{\fg\fl}_1)$ is a trigonometric deformation of the affine Yangian $Y(\widehat{\fg\fl}_1)$.
See \cite{Galakhov:2021vbo} for details.
}
In the classification of quiver BPS algebras the DIM algebra takes a place of a BPS algebra for a $\widehat{\fg\fl}_1$-quiver theory on a space with an extra circle dimension, so that the BPS Hilbert space is described by the K-theory of the target quiver variety, rather than simply cohomologies in the case of quiver Yangians.
The process of Macdonald polynomial degeneration to $r$-Uglov polynomials describing $\widehat{\fg\fl}_r$-theories may be interpreted physically as a phenomenon of (de)construction \cite{Arkani-Hamed:2001kyx}.
(De)construction allows one to mimic KK-modes along a circle dimension of a higher dimensional quiver theory by looping quiver in an infinite node chain instead.

From our perspective it is intriguing to follow how different object of the QFT are transcribed during the (de)construction process of Macdonald polynomials towards $r$-Uglov polynomials.

\subsection{(De)constructing bond factors}\label{sec:deco}

Let us first concentrate  on the structure functions \eqref{bond_factors}  -- entering the definition of the algebra.

For the DIM algebra $T(\widehat{\fg\fl}_1)$ we introduce an exponentiated content function for a box $\Box$ with coordinates $(x_{\Box},y_{\Box})$:
\begin{equation}\label{exponent_box}
	\Omega_{\Box}=q_1^{x_{\Box}}q_2^{y_{\Box}}\,,
\end{equation}
where $q_1$, $q_2$, $q_3$ are deformation parameters of $T(\widehat{\fg\fl}_1)$ subjected to a relation $q_1q_2q_3=1$.

An analog of \eqref{bond_factors} for $T(\widehat{\fg\fl}_1)$ reads:
\begin{equation}
	\Phi\left(\Omega_{\Box},\Omega_{\Box'}\right)=\frac{\left(\Omega_{\Box}-\Omega_{\Box'}q_1\right)\left(\Omega_{\Box}-\Omega_{\Box'}q_2\right)\left(\Omega_{\Box}-\Omega_{\Box'}q_3\right)}{\left(\Omega_{\Box}-\Omega_{\Box'}q_1^{-1}\right)\left(\Omega_{\Box}-\Omega_{\Box'}q_2^{-1}\right)\left(\Omega_{\Box}-\Omega_{\Box'}q_3^{-1}\right)}\,.
\end{equation}

To (de)construct the Macdonald polynomials we calculate limit $\hbar\to 0$ with the following redefinition of parameters:
\begin{equation}\label{deco_limit}
	q_1=e^{\epsilon_1\hbar+\frac{2\pi i}{r}},\quad q_1=e^{\epsilon_2\hbar-\frac{2\pi i}{r}},\quad q_3=e^{-\epsilon_1\hbar-\epsilon_2\hbar}\,.
\end{equation}

Then the box content takes the following form:
\begin{equation}\label{color_shift}
	\Omega_{\Box}=e^{\hbar \omega_\Box}e^{\frac{2 \pi i}{r}c_{\Box}}\,,
\end{equation}
where
\begin{equation}
	\begin{aligned}
		\omega_{\Box}&=\epsilon_1x_{\Box}+\epsilon_2y_{\Box}\mbox{ -- reduced box content,}\\
		c_{\Box}&=(x_{\Box}-y_{\Box})\;{\rm mod}\;r\mbox{ -- box color.}
	\end{aligned}
\end{equation}

In these terms the bond factor reads:
\begin{equation}\label{PhiCol}
	\Phi=\frac{e^{\hbar\left(\omega_{\Box}-\omega_{\Box'}-\epsilon_1\right)}e^{\frac{2\pi i}{r}\left(c_{\Box}-c_{\Box'}-1\right)}-1}{e^{\hbar\left(\omega_{\Box}-\omega_{\Box'}+\epsilon_2\right)}e^{\frac{2\pi i}{r}\left(c_{\Box}-c_{\Box'}-1\right)}-1}
	%%%%%%%%%%%%%%%%
	\frac{e^{\hbar\left(\omega_{\Box}-\omega_{\Box'}-\epsilon_2\right)}e^{\frac{2\pi i}{r}\left(c_{\Box}-c_{\Box'}+1\right)}-1}{e^{\hbar\left(\omega_{\Box}-\omega_{\Box'}+\epsilon_1\right)}e^{\frac{2\pi i}{r}\left(c_{\Box}-c_{\Box'}+1\right)}-1}
	%%%%%%%%%%%%%%%%
	\frac{e^{\hbar\left(\omega_{\Box}-\omega_{\Box'}+\epsilon_1+\epsilon_2\right)}e^{\frac{2\pi i}{r}\left(c_{\Box}-c_{\Box'}\right)}-1}{e^{\hbar\left(\omega_{\Box}-\omega_{\Box'}-\epsilon_1-\epsilon_2\right)}e^{\frac{2\pi i}{r}\left(c_{\Box}-c_{\Box'}\right)}-1}\,.
\end{equation}

Apparently, rather than having a single uniform limit expression \eqref{PhiCol} has a family of limits $\hbar\to 1$ depending on colors $a$ and $b$ of boxes $\Box$ and $\Box'$ respectively
\begin{equation}
	\Phi=\varphi^{(a,b)}\left(\omega_{\Box}-\omega_{\Box'}\right)+O(\hbar)\,,
\end{equation}
where we derive:
\begin{equation}
	\begin{aligned}
		\varphi^{(a,b)}(z)&= 1,\quad \mbox{if }|a-b|>1\,,\\
		\varphi^{(a,a)}(z)&= \frac{z+\epsilon_1+\epsilon_2}{z-\epsilon_1-\epsilon_2}\,,\\
		\varphi^{(a+1,a)}(z)&= \frac{z-\epsilon_1}{z+\epsilon_2}\,,\\
		\varphi^{(a,a+1)}(z)&= \frac{z+\epsilon_1}{z-\epsilon_2}\,,
	\end{aligned}
\end{equation}
where the indices are understood modulo $r$.

\subsection{Yangian times from Macdonald representation}

Next, we repeat the orthogonalization procedure from Appendix \ref{sec:Uglov_def} for the case of Macdonald polynomials
and explain how the former can be embedded into the latter one.

The major discrepancy is the norm on monomials of time variables:
\begin{equation}\label{Mac_norm}
	\left\langle p_{\lambda},p_{\mu}\right\rangle_r=\delta_{\mu,\nu}Z_{\lambda}\times \prod\lm_i\frac{1-q_2^{-\lambda_i}}{1-q_1^{\lambda_i}}\,.
\end{equation}

The orthogonalization procedure discussed in Appendix \ref{sec:Uglov_def} applied with norm \eqref{Mac_norm} directly gives rise to what we might call \emph{unnormalized} Macdonald polynomials.
Define \emph{normalized} Macdonald polynomials ${\rm Mac}_{\lambda}$ in the following way:
\begin{equation}
	{\rm Mac}_{\lambda}={\rm Pol}_{\lambda}\times\prod\lm_{\Box\in\lambda}\left(1-q_1^{{\rm arm}_{\lambda}(\Box)+1}q_2^{-{\rm leg}_{\lambda}(\Box)}\right)\,.
\end{equation}

First, we observe that the norm of Macdonald polynomials:
\begin{equation}
	\left\langle {\rm Mac}_{\lambda},{\rm Max}_{\mu}\right\rangle=\delta_{\lambda,\mu}\CN_{\lambda}
\end{equation}
reads
\begin{equation}
	\CN_{\lambda}=\prod\lm_{\Box\in\lambda}\left(1-q_1^{{\rm arm}_{\lambda}(\Box)+1}q_2^{-{\rm leg}_{\lambda}(\Box)}\right)\left(1-q_1^{{\rm arm}_{\lambda}(\Box)}q_2^{-\left({\rm leg}_{\lambda}(\Box)+1\right)}\right)\,.
\end{equation}

Define operators $e_{k\in\IZ}$, $f_{k\in \IZ}$ of the DIM algebra $T(\widehat{\fg\fl}_1)$ in the Fock module whose vectors are in one-to-one correspondence with the Young diagrams similarly to $Y(\widehat{\fg\fl}_1)$ as we did in Sec. \ref{sec:CryRep}.
These raising/lowering operators modify diagrams by adding/subtracting a box, and the mode number $k$ is defined as a weight for the exponentiated migrating box content \eqref{exponent_box}:
\begin{equation}
	\begin{aligned}
		e_k|\lambda\rangle=&\sum\lm_{\Box\in\lambda^+}{\bf E}_{\lambda,\lambda+\Box}\,\Omega_{\Box}^k\;|\lambda+\Box\rangle\,,\\
		f_k|\lambda\rangle=&\sum\lm_{\Box\in\lambda^-}{\bf F}_{\lambda,\lambda-\Box}\,\Omega_{\Box}^k\;|\lambda-\Box\rangle\,.
	\end{aligned}
\end{equation}

Where the unweighted matrix elements read (cf. \cite{FeiginTsymbaliuk}):
\begin{equation}\label{EF_Mac}
	\begin{aligned}
		{\bf E}_{\lambda,\lambda+\Box}&=\frac{1}{1-q_1}\prod\lm_{\Box'\in\CH_{\lambda}(\Box)}\frac{1-q_1^{{\rm arm}_{\lambda}(\Box')}q_2^{-({\rm leg}_{\lambda}(\Box')+1)}}{1-q_1^{{\rm arm}_{\lambda}(\Box')+1}q_2^{-({\rm leg}_{\lambda}(\Box')+1)}}\prod\lm_{\Box'\in\CV_{\lambda}(\Box)}\frac{1-q_1^{{\rm arm}_{\lambda}(\Box')+1}q_2^{-{\rm leg}_{\lambda}(\Box')}}{1-q_1^{{\rm arm}_{\lambda}(\Box')+1}q_2^{-({\rm leg}_{\lambda}(\Box')+1)}}\,,\\
		{\bf F}_{\lambda,\lambda+\Box}&=(1-q_1)\prod\lm_{\Box'\in\CH_{\lambda}(\Box)}\frac{1-q_1^{{\rm arm}_{\lambda}(\Box')+2}q_2^{-{\rm leg}_{\lambda}(\Box')}}{1-q_1^{{\rm arm}_{\lambda}(\Box')+1}q_2^{-{\rm leg}_{\lambda}(\Box')}}\prod\lm_{\Box'\in\CV_{\lambda}(\Box)}\frac{1-q_1^{{\rm arm}_{\lambda}(\Box')}q_2^{-({\rm leg}_{\lambda}(\Box')+2)}}{1-q_1^{{\rm arm}_{\lambda}(\Box')}q_2^{-({\rm leg}_{\lambda}(\Box')+1)}}\,.
	\end{aligned}
\end{equation}

We would like to construct time operators $\hat \xi_{k\in \IN}$ for this representation compatible with the Macdonald polynomials.
By compatibility we imply the following properties:
\begin{equation}
	\begin{aligned}
		\mbox{a)} & \quad \quad \left[\hat\xi_i,\hat\xi_j\right]=0,\quad \forall\, i,j\,,\\
		\mbox{b)} & \quad \quad \hat\xi_k\cdot {\rm Mac}_{\lambda}(p_1,p_2,\ldots)\Big|_{p_m\to\hat\xi_m}=\left(p_k\,{\rm Mac}_{\lambda}(p_1,p_2,\ldots)\right)\Big|_{p_m\to\hat\xi_m}\,,\\
		\mbox{c)} & \quad \quad {\rm Mac}_{\lambda}(p_1,p_2,\ldots)\Big|_{p_m\to\hat\xi_m}\,|\varnothing\rangle\,=\,|\lambda\rangle\,.
	\end{aligned}
\end{equation}

After calculating several first levels we \emph{conjecture} the following generic form for these operators:
\begin{equation}
	\hat\xi_k|\lambda\rangle=\sum\lm_{\Box_1,\Box_2,\ldots,\Box_k}{\bf E}_{\lambda,\lambda+\Box_1}{\bf E}_{\lambda+\Box_1,\lambda+\Box_1+\Box_2}\ldots{\bf E}_{\lambda+\ldots+\Box_{k-1},\lambda+\ldots+\Box_{k}}\,g_k(\omega_{\Box_1},\omega_{\Box_2},\ldots,\omega_{\Box_k})\,|\lambda+\ldots+\Box_k\rangle\,,
\end{equation}
where form-factors $g_k$ read:
\begin{equation}
	g_1(z_1)=1,\quad g_k(z_1,z_2,z_3,...,z_k)=\frac{\sum\lm_{i=1}^k(-1)^i\left(\!\!\!\begin{array}{c}
			\scriptstyle k-1\\ \scriptstyle i-1
		\end{array}\!\!\!\right)z_i}{\sum\lm_{i=1}^k z_i}\frac{1-q_2^k}{(1-q_2)^k},\mbox{ for }k\geq 2\,,
\end{equation}
where the coefficients in the numerator are ordinary binomial coefficients.

As we have seen in Sec.~\ref{sec:time_op} the first instance when we encounter an issue with the expressability of the time operators in terms of raising operators of $Y(\widehat{\fg\fl}_r)$ occurs at $r=2$ for operator $\hat\xi_2$.
Let us concentrate on this operator, the corresponding form-factor reads in this case:
\begin{equation}\label{g_2}
	g_2(z_1,z_2)=\frac{z_1-z_2}{z_1+z_2}\frac{1-q_2^2}{(1-q_2)^2}\,.
\end{equation}
The (de)construction procedure invokes limit $\hbar\to 0$ with a specific twisted limit \eqref{deco_limit} unlooping the cylinder compact dimension.
Due to this extra twist multipliers in expressions \eqref{EF_Mac} have a coordinate dependent limit $\hbar\to 0$ only if the relative color of the argument is $0\; {\rm mod} \; r$.
This observation gives rise to function $\gamma_r$ defined in \eqref{cyclic} and to reduced expressions \eqref{EF_cyc} for $\bf E$- and $\bf F$-matrix elements.
The rational function in the form-factor \eqref{g_2} behaves differently:
\begin{equation}
	\frac{\Omega_{\Box}-\Omega_{\Box'}}{\Omega_{\Box}+\Omega_{\Box'}}=\frac{1-e^{\frac{2\pi i}{r}\left(c_{\Box'}-c_{\Box}\right)+\hbar\left(\omega_{\Box'}-\omega_{\Box}\right)}}{1+e^{\frac{2\pi i}{r}\left(c_{\Box'}-c_{\Box}\right)+\hbar\left(\omega_{\Box'}-\omega_{\Box}\right)}}\sim\left\{\begin{array}{ll}
		\omega_{\Box}-\omega_{\Box'}, & r=1\,,\\
		\dfrac{1}{\omega_{\Box}-\omega_{\Box'}}, & r=2\,.
	\end{array}\right.
\end{equation}
In the case $r=1$ when there is no color dependence the twist term vanishes, and the expression reduces to a simple zero producing the commutator term $[e_1,e_0]$ as in \eqref{beta=1_xi_2}.
Whereas in the case $r=2$ the colors of boxes differ by 1, and the twist term produces a minus sign shifting the zero from the numerator to the denominator -- this is a pole of \eqref{xi-times}.

The apparent asymmetry between $q_1$ and $q_2$ in the formula (135), as well as monotonic increase of the pole degree
with $k$ remains an open question, which requires better understanding and an appropriate interpretation. 

%Comments on $q_2$ dependence etc ???

\subsection{On other versions of DIM}

At the end of this Appendix we would like to pose an important question for further research.

As we have just seen, $r$-Uglov polynomials $U^{(r)}$ are naturally embedded into Macdonald ones --
which in turn are a typical representation of $T(\widehat{\fg\fl}_1)$.

As we explained in this paper, $r$-Uglov polynomials provide a nice representation of Yangian $Y(\widehat{\fg\fl}_r)$.

The question is, if there is an even more natural embedding of $U^{(r)}$ into representation of
$T(\widehat{\fg\fl}_r)$, not just $T(\widehat{\fg\fl}_1)$?
There is no contradiction -- both embeddings can exist and {\it co}exist,
but the form of this coexistence can appear interesting and practically useful.

Additionally, this question may be posed in an alternative way: whether there exists a natural system of orthogonal polynomials describing Fock modules of $T(\widehat{\fg\fl}_r)$?
Moreover if this is the case, are we allowed to extend this construction and obtain these -- as far as we are aware -- potentially \emph{new} systems of orthogonal polynomials for $T(\widehat{\fg\fl}_r)$ as some limits of analogous constructions in the elliptic deformation $E(\widehat{\fg\fl}_1)$ of the affine Yangian?
In other words, if this construction survives adding an extra loop to the algebra.

\bibliographystyle{utphys}
\bibliography{biblio}

\end{document}